\documentclass[onecolumn]{article}

\usepackage[utf8]{inputenc}
\usepackage[big]{dgruyter}

\usepackage{bbm}
\usepackage{subcaption}
\usepackage{amsthm}
\usepackage{amsfonts}
\usepackage{multirow}
\usepackage{multicol}
\usepackage{appendix}

\newtheorem{assumption}{Assumption}

\newtheorem{proposition}{Proposition}
\newtheorem{lemma}{Lemma}

\newcommand\independent{\protect\mathpalette{\protect\independenT}{\perp}}
\def\independenT#1#2{\mathrel{\rlap{$#1#2$}\mkern2mu{#1#2}}}

\begin{document}
  \articletype{Research Article{\hfill}Open Access}

  \author*[1]{Richard A.~J.~Post}

\author[2]{Edwin R.~van den Heuvel}

  \affil[1]{Department of Epidemiology, Erasmus Medical Center, Rotterdam, the Netherlands, e-mail: r.a.j.post@erasmusmc.nl}
  \affil[2]{Department of Mathematics and Computer Science, Eindhoven University of Technology, Eindhoven, the Netherlands}
  
  \title{\huge Beyond Conditional Averages: Estimating The Individual Causal Effect Distribution} 

  \runningtitle{Beyond Conditional Averages}


  \begin{abstract}
{In recent years, the field of causal inference from observational data has emerged rapidly. The literature has focused on (conditional) average causal effect estimation. When (remaining) variability of individual causal effects (ICEs) is considerable, average effects may be uninformative for an individual. The fundamental problem of causal inference precludes estimating the joint distribution of potential outcomes without making assumptions. In this work, we show that the ICE distribution is identifiable under (conditional) independence of the individual effect and the potential outcome under no exposure, in addition to the common assumptions of consistency, positivity, and conditional exchangeability. Moreover, we present a family of flexible latent variable models that can be used to study individual effect modification and estimate the ICE distribution from cross-sectional data. How such latent variable models can be applied and validated in practice is illustrated in a case study on the effect of Hepatic Steatosis on a clinical precursor to heart failure. Under the assumptions presented, we estimate that $20.6 \%$ ($95\%$ Bayesian credible interval: $8.9 \%, 33.6 \%$) of the population has a harmful effect greater than twice the average causal effect.}
\end{abstract}
  \keywords{Causal inference, Heterogeneity of treatment effects, Precision medicine, Bayesian analysis, Random effects models.}
   \classification[MSC]{62D20, 62F15, 62P10}

  \journalname{Journal of Causal Inference}
\DOI{XXX}
  \startpage{1}
  \received{February 1, 2024}
  \accepted{April 2, 2025}

  \journalyear{2025}
  \journalvolume{X}

\maketitle
\section{Introduction}

\begin{figure}[t]
	\centering
	\begin{subfigure}{.35\textwidth}
		\resizebox{1\linewidth}{!}{\includegraphics{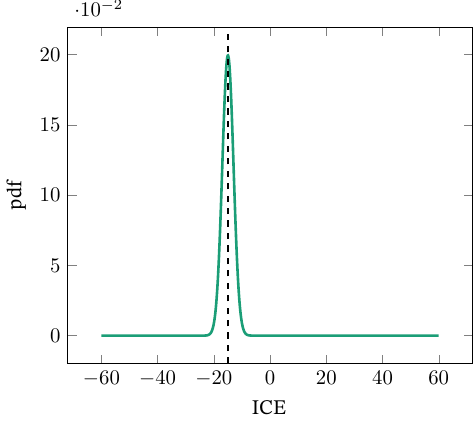}}
		\caption{}\label{CH310c}	
	\end{subfigure}  \hspace{0.1cm}
	\begin{subfigure}{.35\textwidth}
		\resizebox{1\textwidth}{!}{\includegraphics{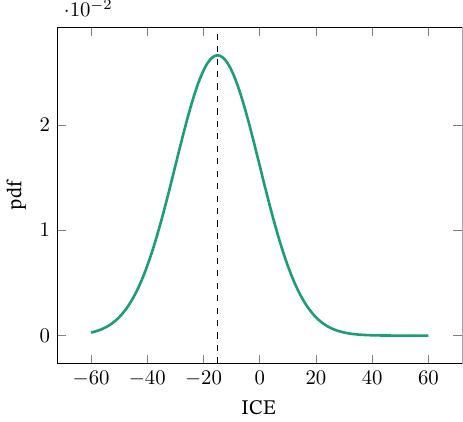}}
		\caption{}\label{CH310a}	
	\end{subfigure}
	\begin{subfigure}{.35\textwidth}
		\resizebox{1\textwidth}{!}{\includegraphics{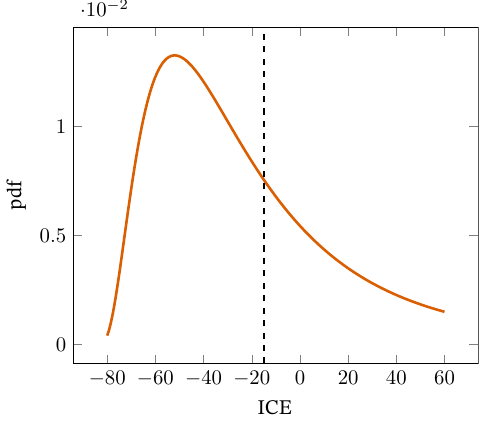}}
		\caption{}\label{CH310d}	
	\end{subfigure} \hspace{0.1cm}
	\begin{subfigure}{.35\textwidth}
		\resizebox{1\textwidth}{!}{\includegraphics{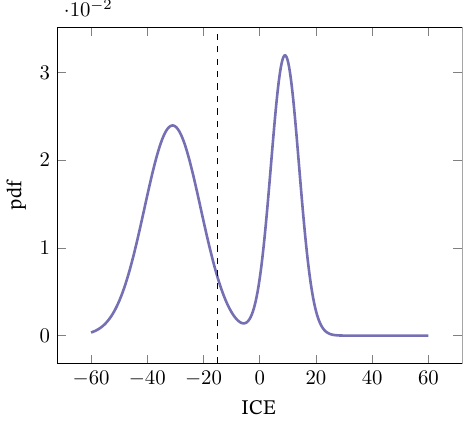}}
		\caption{}\label{CH310f}	
	\end{subfigure}
	
	\caption{Examples of effect distributions where the (conditional) ATE equals $-15$. ICEs follow a Gaussian distribution with $\sigma=2$ (a) and $\sigma{=}15$ (b) respectively, a log-normal distribution ($\mu{=}4, \sigma{=}\sqrt{0.5}),$ c), or a Gaussian mixture distribution with $p{=}0.6, \mu_{1}{=}-31, \sigma_{1}{=}10, \mu_{2}{=}9$ and $\sigma_{2}{=}5$ (d). }\label{CH3FIG1}
\end{figure}

The main result of an epidemiological study is often summarized by an average treatment effect (ATE). Consequently, the ATE might (subconsciously) be interpreted as the causal effect for each individual, while individuals may react differently to exposures. These individual causal effects (ICEs) can be highly variable and may even have opposite signs \citep{Hand1992, Greenland2019}. The effect distribution within a (sub)population might be skewed, high in variability, or multimodal. Examples of ICE distributions for which the ATE is equal to $-15$ are presented in Figure \ref{CH3FIG1}.  One would like to go beyond mean estimation and draw inferences on the distribution of the ICEs. This inference involves the joint distribution of the potential outcomes of an individual that is not identifiable since the pair of potential outcomes under different levels of intervention cannot be observed simultaneously \citep{Holland1986}. This fundamental problem of causal inference makes it impossible to estimate ICEs without making strong assumptions. Marginal causal effects can be characterized as generic functionals of the densities of potential outcomes \cite{Kennedy2023}. Moreover, a difference in the shape of the potential outcome distributions can inform on some properties of the ICE distribution. For example, as we will explain further in this work, effect homogeneity can be ruled out when the variances of exposed and unexposed individuals are different in a randomized controlled trial (RCT). Under additional assumptions, the ICE distribution becomes identifiable and can be learned from the difference in shapes of the potential outcome distributions.  

Nowadays, the variability in ICEs is typically studied by estimating conditional average treatment effects (CATEs), which equals the mean of the conditional ICE distribution given certain observed features. Standard causal inference methods can typically include covariates to account for such effect modification, e.g.~strata-specific marginal structural models \citep{Robins2000}, and many machine learning (ML) algorithms have been proposed for flexible CATE estimation (see \citet{Caron2022} for a detailed review). In this ML literature, a CATE based on many observed features is sometimes considered equivalent to the ICE \citep{Lu2018}. In the ideal case of almost homogeneous conditional causal effects, such as for a subpopulation with the effect distribution shown in Figure \ref{CH310c}, the CATE can be used to make treatment decisions for an individual. However, the CATE may not be informative enough when given the levels of the measured features the ICEs among the subpopulation are still highly variable, as in figures \ref{CH310a} and \ref{CH310d}. Despite accounting for many features of the patient, other and unknown features might modify the exposure effect so that in the subpopulation the exposure is expected to harm some while others benefit (e.g.~Figure \ref{CH310f}). Despite the increasing number of covariates collected in epidemiological studies, significant effect modifiers may remain unmeasured so that the personalized CATE deviates from the personal ICE \citep{Post2024}. Therefore, (remaining) effect heterogeneity should be taken into account for decision-making \citep{Kallus2022, Ben-Michael2024, Li2023, Mueller2023, Sarvet2023}.

 In recent work, we have explained how CATE estimation using machine learning methods may be extended to additionally estimate the remaining conditional variance of the ICE assuming conditional independence of 
 the individual effect and the potential outcome under no exposure, a causal assumption \citep{Post2024}. The conditional ICE distribution is only identified by this conditional variance and the CATE when assuming (conditionally) Gaussian-distributed effects. Under the conditional independence assumption, the Gaussianity of the effect is not a causal assumption because it could be verified with sufficient observed data. In this work, we shift focus from the (conditional) ICE variance to the (conditional) ICE distribution without assuming the shape of the effect distribution. We will explain under what assumptions the (conditional) ICE distribution becomes identifiable so that latent variable models may be used to estimate the distribution. We will present a semi-parametric linear mixed model to estimate the (conditional) ICE distribution when the assumptions apply and discuss the complications in estimation due to the heterogeneity of the effect of confounders on the outcome. 

Firstly, we will introduce our notation and framework to formalize individual effect heterogeneity in Section \ref{CH3sec2}. In Section \ref{CH3sec2H}, we will review the literature related to ICE and illustrate how and under which assumptions the ICE distribution can be identified from an RCT and observational data, respectively. Subsequently, the semi-parametric causal mixed models and possible methods to fit these models are presented in Section \ref{CH3sec5}. In Section \ref{CH3sec6}, the Framingham Heart Study is considered to illustrate how the reasoning presented in this work could be used in practice. The modeling considerations are discussed in detail. Finally, we reflect on the work's importance, limitations, and future research in Section \ref{CH3sec7}. 

\section{Notation and framework} \label{CH3sec2}
Probability distributions of factual and counterfactual outcomes are defined in the potential outcome framework \citep{Neyman1990, Rubin1974}. Let $Y_{i}$ and $A_{i}$ represent the (factual) stochastic outcome and the random exposure assignment level of individual $i$. Let $Y^{a}_{i}$ equal the potential outcome (counterfactual when $A_{i} \neq a$) under an intervention on the exposure to level $a$. We thus assume deterministic potential outcomes so that each level of exposure corresponds to a specific outcome for each individual (but its value typically differs between individuals) \citep{Robins1989, VanderWeele2012}.

We will consider only two exposure levels, $a \in \{0,1\}$, with $0$ indicating no exposure. The individual causal effect of an arbitrary individual $i$ from the population of interest is defined on the additive scale as $Y^{1}-Y^{0}$ \citep[Chapter 1]{Hernan2019}. 

Throughout this work, we will assume causal consistency, implying that potential outcomes are independent of the assigned exposure levels of other individuals (no interference) and that there are no different versions of the exposure levels \citep{Cole2009}.  \begin{assumption}{\textbf{Causal consistency}}\label{CH2A2}
	$$Y_{i}~{=}~Y_{i}^{A_{i}}$$
\end{assumption} This definition of causal consistency is also referred to as the stable unit treatment value assumption \citep[Assumption 1.1]{Imbens2015}.
Furthermore, we will assume that for all levels of measured features $\boldsymbol{X}$, the probability of exposure is bounded away from $0$ and $1$, known as positivity \citep[Chapter 3]{Hernan2019}. \begin{assumption}{\textbf{Positivity}}\label{CH2A3} $$
\forall \boldsymbol{x}{:}~ 0 < \mathbb{P}(A~{=}~1  \mid \boldsymbol{X}{=}\boldsymbol{x}) < 1 $$\end{assumption} 

Next, we will introduce a structural model and random variables necessary to formalize the relevant causal relations and joint distribution of potential outcomes. The variability in exposure assignment $A$ among individuals is accommodated by the variable $N_{A}$. Individuals can have different potential outcomes under no exposure, $Y^{0}$, represented by a population average $\theta$ and an individual specific deviation $N_{Y}$. The latter random variable results in dependence between $Y^{0}$ and $Y^{1}$. In addition, ICE is equal to the population average effect $\tau$ and an individual random effect deviation $U_{1}$. We describe the cause-effect relations with a structural causal model (SCM), which is commonly used in the causal graphical literature, see e.g.~\citet[Chapter 1.4]{Pearl2009book} and ~\citet[Chapter 6]{Peters2017}, to formalize the system behind the observed outcomes. Instead, we include both individual effect modifiers $U_{1}$ and the latent common cause of potential outcomes $N_{Y}$ to describe the two potential outcomes of an individual jointly. A SCM as presented in this paper is therefore a union of the SCM for observations ($A{=}a$), and the so-called intervened SCMs for all possible $do(A{=}a)$. It consists of a joint probability distribution for $(N_{A}, N_{Y}, U_{1})$ and a structural assignment $f_{A}$\footnote{When $N_{A}$ follows a Uniform [0,1] distribution, $f_{A}(\cdot)$ equals the quantile function of $A$.} such that
\begin{center}
\fbox{%
\parbox{0.9\linewidth}{%
\begin{align}\label{CH3SCMsurv2}
		A_{i}&~{:=}~f_{A}(N_{Ai})   \\\nonumber
		Y^{a}_{i} &~{:=}~ \theta + N_{Yi} + \left(\tau+U_{1i}\right)a. 
		\end{align}}}\\ 
\end{center} Note that $U_{1i}$ can still depend on $N_{Yi}$, so that SCM \eqref{CH3SCMsurv2} is a saturated model and does not impose any restrictions. For example, consider a system where $Y_{i}^{1}= 2Y_{i}^{0} = 2\theta + 2N_{Yi}$, then $\tau = \theta$ and $U_{1i}=N_{Yi}$. Although no other observed features $\boldsymbol{X}$ are present in the SCM \eqref{CH3SCMsurv2}, $U_{1i}$ and $N_{Yi}$ can depend on $\boldsymbol{X}$. Later in this work, the SCM \eqref{CH3SCMsurv2} will be reparamaterized by including observed features $\boldsymbol{X}$. The joint distribution of $U_{1i}$ and $N_{Yi}$ is not identifiable without additional assumptions and will play a vital role in this work. The SCM \eqref{CH3SCMsurv2} also describes the data-generating mechanism since $Y_{i}^{A_{i}}=Y_{i}$. 

There exists confounding when $N_{Ai} \not\independent N_{Yi}, U_{1i}$, so that $A \not\independent Y^{0}, Y^{1}$. For observational data, many causal inference methods require the absence of unmeasured confounding, i.e., observed features $\boldsymbol{X}$ should contain a sufficient adjustment set, and the observed features should not contain colliders \citep{Munafo2017}. Then, $Y^{0}, Y^{1} \independent A \mid \boldsymbol{X}$, referred to as conditional exchangeability \citep[Chapter 3]{Hernan2019}, and the marginal distributions of the potential outcomes are identified by the observed conditional distributions. 
\begin{assumption}{\textbf{Conditional exchangeability}}\label{CH2A1}
	$$Y^{0}, Y^{1} \independent A \mid \boldsymbol{X}$$
\end{assumption}

As discussed in the introduction, causal effect heterogeneity is often addressed by studying the CATEs in subpopulations defined by measured features $\boldsymbol{X}$ (e.g.,~sex, age or co-morbidity) that may be effect modifiers so that $Y_{1}-Y_{0} \not\independent\boldsymbol{X}$. The CATE estimand equals $\mathbb{E}[Y^1-Y^0 \mid \boldsymbol{X}=\boldsymbol{x}]$, so that with the parameterization in SCM \eqref{CH3SCMsurv2}, the CATE for $\boldsymbol{X}=\boldsymbol{x}$ is equal to $\tau + \mathbb{E}[U_{1} \mid \boldsymbol{X}=\boldsymbol{x}]$. Under assumptions \ref{CH2A2}, \ref{CH2A3} and \ref{CH2A1} the CATE is identifiable and using causal methods such as inverse probability of treatment weighting (IPTW) for adjustment, unbiased CATE estimates can be obtained \citep[Chapter 2]{Hernan2019}. Note that conditioning on $\boldsymbol{X}$ can thus have two purposes: adjusting for confounding to obtain conditional exchangeability or explaining part of the effect heterogeneity. For example, when studying effect heterogeneity from a randomized experiment, one does not have to adjust for $\boldsymbol{X}$ for exchangeability to hold but can condition to study the CATEs for different levels of $\boldsymbol{X}$. 

The CATE equals the mean of the conditional ICE distribution, $\mathbb{P}(Y^1-Y^0 \leq y \mid \boldsymbol{X}=\boldsymbol{x})$. The ICE distribution is the estimand of interest in this work. In contrast to the CATE, the ICE distribution itself can only be identified from cross-sectional data after assuming how $Y^{1}-Y^{0}$ depends on $Y^{0}$. We will show in Section \ref{CH3sec2H} that a sufficient assumption for identifiability of the ICE distribution (when assumptions \ref{CH2A2}, \ref{CH2A3}, and \ref{CH2A1} apply) would be conditional independence of the ICE and $Y^{0}$.
 \begin{assumption}{\textbf{Conditional independent effect deviation}}\label{CH2A4}
	$$ Y^{1}-Y^{0} \independent Y^{0}  \mid \boldsymbol{X}{=}\boldsymbol{x} $$\end{assumption}
 
For the parameterization of the cause-effect relations in terms of SCM \eqref{CH3SCMsurv2}, this is equivalent to assuming $U_{1} \independent N_{Y} \mid \boldsymbol{X}=\boldsymbol{x}$. Assumption \ref{CH2A4} means that after taking into account features $\boldsymbol{X}$, the potential outcome under no exposure is not informative for the value of the individual effect. The assumption could, for example, apply when the causal effect of a treatment is the result of a catalytic reaction mechanism. Then, the individual causal effect of exposure depends on the amount of a catalyst present for an individual, while conditionally, on features $\boldsymbol{X}$, this amount of catalyst is not associated with $Y^{0}$. In \cite{Post2024}, we present an example of the effect of an antiplatelet medicine, which depends on the conversion to an active metabolite accomplished by a specific enzyme. Heterogeneity in ICE results from heterogeneity in the amount of enzyme present. After adjusting for features that are related to the amount of enzyme, the actual amount of enzyme does not inform on $Y^0$, so that $Y^{1}-Y^{0} \independent Y^{0} \mid \boldsymbol{X}$. 

Realize that Assumption \ref{CH2A4} is violated when there exists a feature that, despite conditioning on $\boldsymbol{X}$, affects both $Y^{1}-Y^{0}$ and $Y^0$ and is not adjusted for. Moreover, it is good to realize that when Assumption \ref{CH2A4} applies, then $\frac{Y^{1}}{Y^0}$ will generally not be independent of $Y^{0}$ given $\boldsymbol{X}$. In our work, we are interested in the causal effect defined as the difference of potential outcomes, and thus Assumption \ref{CH2A4} involves that difference. When the focus is on another contrast of potential outcomes, an assumption similar to \ref{CH2A4} but involving that contrast will be required.

\section{Identifiability of the ICE distribution}\label{CH3sec2H}
As a result of the fundamental problem of causal inference, studying properties of the conditional ICE distribution other than the CATE (which equals the mean) received less attention in the literature. For binary outcomes $Y$ and exposures $A$, the ICE distribution can be described by the probability of necessity (PN, i.e., $\mathbb{P}(Y^0 = 0 | A=1, Y=1)$), the probability of sufficiency (PS, i.e.,  $\mathbb{P}(Y^1 = 1 | A=0, Y=0)$) and the probability of necessity and sufficiency (PNS, i.e.,  $\mathbb{P}(Y^1 = 1, Y^0 = 0)$) that is also referred to the probability of benefit\citep{Pearl1999}. For binary exposures and outcomes, the PN is identifiable under a monotonic effect assumption ($Y^1 \geq Y^0$) \citep{Pearl2009book}. For the case with multiple causes that could affect each other, similar monotonicity assumptions have been presented to identify the posterior total and direct causal effects \citep{Lu2022}. Furthermore, again for binary exposures, under the assumption of monotonic incremental treatment effect (i.e, $0 \leq Y^{1}-Y^{0} \leq 1$), the PS has been shown to be identifiable for ordinal outcomes \citep{Zhang2024}. Assuming the absence of unmeasured confounding, bounds on the PN, PS and PNS obtained from combined experimental and observational data have been presented \citep{Pearl1999}, and their applicability in practice was shown \citep{Tian2000}. Similarly, to study the conditional (on covariates) ICE distribution, conditional bounds of the PNS can be estimated \citep{Mueller2023}. When combining experimental and observational data, bounds have also been presented for multi-valued outcomes and exposures \citep{Li2022}.  Moreover, for multi-valued outcomes, binary exposures, and when only using experimental data, a consistent estimator for the bounds of the treatment benefit ratio (TBR) \citep{Huang2017} and a method to estimate corresponding confidence intervals have been presented \citep{Huang2019}. The TBR is equal to $\mathbb{P}(Y^1-Y^0>0)$ when an increase in the outcome is beneficial and equal to $\mathbb{P}(Y^1-Y^0<0)$ when a decrease is beneficial. Bounds have also been presented for the treatment harming rate (THR, $\mathbb{P}(Y^1-Y^0<0)$ when an increase in the outcome is beneficial), also known as the fraction negatively affected \citep{Gadbury2004, Zhang2013}. 
These bounds can be improved by restricting the range of the Pearson correlation coefficient between potential outcomes\citep{Wu2024}. Recently, a method to estimate lower confidence limits for quantiles of the ICE distribution from experimental data has been proposed for general outcomes and binary exposures, allowing inference on the proportion of units with ICEs passing any threshold \citep{Su2023}. \citet{Su2023} have also presented a sensitivity analysis to apply the methodology to observational data and quantify the effect of unmeasured confounding. Furthermore, ideas from conformal inference have been leveraged to derive individualized prediction intervals for ICEs with $1- \alpha$ marginal coverage guarantees assuming the absence of unmeasured confounding \citep{Lei2020}. More specifically, \citet{Lei2020} present conformalized counterfactual inference to derive prediction intervals for both potential outcomes given the observed features. Moreover, they present both exact and inexact methods to construct intervals for the conditional ICE and demonstrate with numerical experiments that the exact methods are conservative. To robustify the method in cases where unmeasured confounding might exist, sensitivity analyses have been proposed \citep{Jin2022, Yin2022}. Finally, extensions focus on conditional rather than marginal coverage guarantees \citep{Chernozhukov2023, Chernozhukov2021}. 

Besides the nonparametric approaches, non-verifiable assumptions on the joint distribution of the potential outcomes $(Y^0, Y^1)$ have been made by using latent variable models to estimate the TBR \citep{Zhang2013, Yin2018, Laubender2020} and the ICE distribution \citep{Shahn2017}. There exists work in which it is assumed that $Y^1 \independent Y^0$ given features $\boldsymbol{X}{=}\boldsymbol{x}$, while $Y^1-Y^0$ given $\boldsymbol{X}{=}\boldsymbol{x}$ is assumed to follow a Gaussian mixture distribution \citep{Shahn2017} or is degenerate \citep{Zhang2013}. \citet{Yin2018} assume Assumption \ref{CH2A4} but restrict the conditional ICE distribution given $\boldsymbol{X}{=}\boldsymbol{x}$ to be Gaussian.  Instead of a direct assumption on the joint distribution of the potential outcomes, \citet{Laubender2020} assume that next to $Y$, a biomarker is observed that is a surrogate of $Y^1+Y^0$ and by assuming that $Y^0$, $Y^1$ and the biomarker follow a multivariate Gaussian distribution the conditional TBR becomes identifiable. 

In this section, we discuss the non-identifiability of the ICE as a result of the fundamental problem in more detail and explain why Assumption \ref{CH2A4} can result in identifiability. To build intuition, we begin by showing how the fundamental problem of causal inference results in the non-identifiability of the conditional variance. 

\subsection{Conditional variance}
In the absence of remaining effect heterogeneity in subpopulations defined by levels of $\boldsymbol{X}$, the conditional distribution of $Y^{1} - Y^{0} {\mid} \boldsymbol{X}{=}\boldsymbol{x}$ is degenerate in the CATE. Effect heterogeneity is then completely described by modification of the mean effect per subpopulation defined by $\boldsymbol{X}$ and results in equal conditional variance for observed outcomes among exposed and non-exposed individuals. 
\begin{proposition}\label{prop1}
If Assumptions \ref{CH2A2},  \ref{CH2A3} and \ref{CH2A1} apply, and $Y^1-Y^0 \mid \boldsymbol{X}=\boldsymbol{x}$ is degenerate, then 
\begin{equation}
\text{var}(Y\mid A{=}1, \boldsymbol{X}{=}\boldsymbol{x} ) = \text{var}( Y\mid A{=}0,\boldsymbol{X}{=}\boldsymbol{x}).
\end{equation}
\end{proposition}
 The result of Proposition \ref{prop1} concerns observational quantities and can thus be verified when sufficient data are available without making additional assumptions. In that case, it is also possible that the equality in conditional variances does not hold, which by the contrapositive of Proposition \ref{prop1} is equivalent to the existence of remaining effect heterogeneity. In particular, if the conditional variance among treated individuals is larger than non-treated individuals, then the conditional ICE distribution is non-degenerate, as shown in Proposition \ref{prop2}. 
 \begin{proposition}\label{prop2}
If Assumptions \ref{CH2A2}, \ref{CH2A3} and \ref{CH2A1} apply, and 
\begin{equation}
\text{var}(Y\mid A{=}1, \boldsymbol{X}{=}\boldsymbol{x} ) > \text{var}( Y\mid A{=}0,\boldsymbol{X}{=}\boldsymbol{x}),
\end{equation}
then $Y^{1}-Y^{0} \mid \boldsymbol{X}=\boldsymbol{x}$ is non-degenerate. 
\end{proposition}
For an exposure that would reduce the heterogeneity in the outcomes, i.e.,~$\text{var}(Y^1, \boldsymbol{X}{=}\boldsymbol{x} ) < \text{var}( Y^0,\boldsymbol{X}{=}\boldsymbol{x})$, there is remaining heterogeneity in the absence of exposure and $A=a$ could be relabelled as $\tilde{A}=(1-a)$. When there exists $\boldsymbol{x}$ for which the condition in Proposition \ref{prop2} holds, there should be remaining effect heterogeneity, but the conditional ICE distribution is not identifiable due to the fundamental problem of causal inference. For example, the conditional variance 
\begin{equation}
	\text{var}(Y^{1}-Y^{0}{\mid}\boldsymbol{X}{=}\boldsymbol{x}) = \text{var}(Y^{1}{\mid}\boldsymbol{X}{=}\boldsymbol{x}) + \text{var}(Y^{0}{\mid}\boldsymbol{X}{=}\boldsymbol{x})-2 \text{cov}(Y^{1}, Y^{0} {\mid}\boldsymbol{X}{=}\boldsymbol{x}),
\end{equation} is not identifiable without making additional assumptions.\footnote{Without assumptions, only a lower bound on the variance of the ICE follows from the Cauchy-Schwarz inequality and equals
\begin{equation*}
\text{var}(Y^{1}-Y^{0}{\mid}\boldsymbol{X}{=}\boldsymbol{x}) \geq \text{var}(Y^{1}{\mid}\boldsymbol{X}{=}\boldsymbol{x}) + \text{var}(Y^{0}{\mid}\boldsymbol{X}{=}\boldsymbol{x}) -2\sqrt{\text{var}(Y^{1}{\mid}\boldsymbol{X}{=}\boldsymbol{x}) \text{var}(Y^{0}{\mid}\boldsymbol{X}{=}\boldsymbol{x})}.
\end{equation*}} The converse of proposition \ref{prop1} and \ref{prop2} are false since, in terms of SCM \eqref{CH3SCMsurv2}, $\text{var}(Y^{1} \mid \boldsymbol{X}{=}\boldsymbol{x}) = \text{var}(Y^{0} \mid \boldsymbol{X}{=}\boldsymbol{x}) + \text{var}(U_{1} \mid \boldsymbol{X}{=}\boldsymbol{x}) + 2\text{cov}(U_{1}, Y^{0} \mid \boldsymbol{X}{=}\boldsymbol{x})$, so that $\text{var}(Y^{1} \mid \boldsymbol{X}{=}\boldsymbol{x})$ can equal $\text{var}(Y^{0} \mid \boldsymbol{X}{=}\boldsymbol{x})$ when the conditional effect distribution is degenerate or non-degenerate.  

If the conditional covariance, $\text{cov}(Y^1,Y^0 \mid \boldsymbol{X}=\boldsymbol{x})=\mathbb{E}[Y^1Y^0 \mid \boldsymbol{X}{=}\boldsymbol{x}]-\mathbb{E}[Y^1\mid \boldsymbol{X}{=}\boldsymbol{x}]\mathbb{E}[Y^0\mid \boldsymbol{X}{=}\boldsymbol{x}]$, could be expressed in terms of $\text{var}(Y^a \mid \boldsymbol{X}{=}\boldsymbol{x})$, e.g., when Assumption \ref{CH2A4} applies, the conditional ICE variance becomes identifiable \citep{Post2024}. 

\subsection{Conditional ICE distribution}
Understanding the conditional variance is a first step beyond the CATEs when interested in the conditional ICE distribution. We continue by expressing the ICE distribution in terms of the observed outcome distribution given $A$ and $\boldsymbol{X}$, i.e.~$F_{Y \mid A, \boldsymbol{X}}$. 

The conditional distribution of $Y^{1}$ or $Y^{0}$ is commonly expressed in terms of $F_{Y\mid A, \boldsymbol{X}}$ using the g-formula \citep[Chapter 13]{Hernan2019}. By conditional exchangeability, $Y^{a}{\mid}\boldsymbol{X}{=}\boldsymbol{x}$ is equal in distribution to $Y^{a}{\mid} \boldsymbol{X}{=}\boldsymbol{x}, A{=}a$ which, by causal consistency, is equal in distribution to $Y{\mid}\boldsymbol{X}{=}\boldsymbol{x}, A{=}a$. Thus, the conditional distribution of the potential outcomes can be estimated from the observed data. However, since $\{A_{i}=1\}$ and $\{A_{i}=0\}$ are mutually exclusive, the g-formula reasoning does not suffice to express the joint distribution of the potential outcome in terms of $F_{Y}$. It is only after conditioning on the latent shared cause, $N_{Y}$ in SCM \eqref{CH3SCMsurv2}, that the potential outcomes of an individual become independent \citep{Zhang2013}, i.e., 
$Y^{1} \independent Y^{0} ~{\mid}~ \boldsymbol{X}, N_{Y}~$ (since  $Y^{0} {\mid} \boldsymbol{X}, N_{Y}$ is degenerate) while  $Y^{1}\not \independent Y^{0} ~{\mid}~ \boldsymbol{X}$. After conditioning on $N_{Y}$, the joint distribution could be factorized in two parts, and for each piece, it is possible to condition on either $\{A{=}1\}$ or $\{A{=}0\}$ respectively. For the SCM \eqref{CH3SCMsurv2}, the ICE distribution can be expressed in the distributions of the observed outcomes as presented in Lemma \ref{CH5th2}. 

\begin{lemma}\label{CH5th2}
	If assumptions \ref{CH2A2},  \ref{CH2A3} and \ref{CH2A1} apply, and the cause-effect relations are parameterized in terms of SCM \eqref{CH3SCMsurv2}, then $\mathbb{P}(Y^{1}-Y^{0} \leq y ~{\mid}~ \boldsymbol{X}{=}\boldsymbol{x})$ is equal to
\begin{equation}\label{CH3Ffinal}
\int_{-\infty}^{\infty}\int_{y_{1}=-\infty}^{\infty} \int_{y_{2}=y_{1}-y}^{\infty}  1 dF_{Y\mid  A{=}1, N_{Y}{=}n_{Y}, \boldsymbol{X}{=}\boldsymbol{x}}(y_{1}) dF_{Y\mid A{=}0,  N_{Y}{=}n_{Y}, \boldsymbol{X}{=}\boldsymbol{x}}(y_{2}) dF_{N_{Y} \mid \boldsymbol{X}{=}\boldsymbol{x}}(n_{Y}),
\end{equation}
where $\int g(x) dF_{X}(x)$ is the Lebesgue-Stieltjes integral of $g(X)$ with respect to probability law $F_{X}$ and $Y$ is a degenerate random variable given $A=0, N_{Y}=n_{Y}$, and $\boldsymbol{X}{=}\boldsymbol{x}$. The variables of integration $n_{Y}$, $y_1$ and $y_{2}$ are integrated over the support of $N_{Y}$, $Y^{1}$ and those values in the support of $Y^{0}$ that are larger than $y_{1} - y$ respectively. 
\end{lemma}

Inference on the ICE distribution thus requires inference on the conditional distribution of $Y \mid A=0, \boldsymbol{X}{=}\boldsymbol{x}$, characterized by  $\mathbb{E}[Y~{\mid}~A{=}0, \boldsymbol{X}{=}\boldsymbol{x}]$ and the remaining variability of $N_{Y}$ given $\boldsymbol{X}$,  $F_{N_{Y}{\mid}\boldsymbol{X}{=}\boldsymbol{x}}$, and is identifiable under assumptions \ref{CH2A2}, \ref{CH2A3} and \ref{CH2A1}.  Moreover, the conditional expectation, $\mathbb{E}[Y~{\mid}~A{=}1, \boldsymbol{X}{=}\boldsymbol{x}]$, of the exposed individuals is required and identifiable. Although the distribution of $F_{Y{\mid}A{=}1, \boldsymbol{X}{=}\boldsymbol{x}}$ is identifiable, the conditional distribution $F_{Y\mid  A{=}1, N_{Y}{=}n_{Y}, \boldsymbol{X}{=}\boldsymbol{x}}$
is not as we cannot distinguish between $U_{1}$ and $N_{Y}$ due to the fundamental problem of causal inference. The resulting expression in Lemma \ref{CH5th2} is thus not identifiable without an additional assumption on the conditional distribution of $N_{Y}$ and $U_{1}$. If Assumption \ref{CH2A4} applies, the ICE distribution becomes identifiable and the estimation of the conditional ICE distribution becomes equivalent to the statistical problem of (conditional) density deconvolution \citep{Meister2009} as shown in Proposition \ref{Cor1}. 
\begin{proposition}\label{Cor1}
Let the random variable $G_{0}$ be equal in distribution to $Y\mid A{=}0, \boldsymbol{X}{=}\boldsymbol{x}$ and let $G_{1}$ be a random variable so that $G_{1} \independent G_{0}$ and $G_{0}+G_{1}$ is equal in distribution to $Y\mid A{=}1, \boldsymbol{X}{=}\boldsymbol{x}$. If assumptions \ref{CH2A2}, \ref{CH2A3}, \ref{CH2A1} and \ref{CH2A4} apply, then $
\mathbb{P}(Y^{1}-Y^{0} \leq y ~{\mid}~ \boldsymbol{X}{=}\boldsymbol{x}) = \mathbb{P}(G_{1} \leq y)$. 
 \end{proposition} 
 In words, if Proposition \ref{Cor1} applies, $Y^1-Y^0 \mid \boldsymbol{X}=\boldsymbol{x}$ is equal in distribution to the distribution of the difference between independent realizations from $Y\mid A{=}1, \boldsymbol{X}{=}\boldsymbol{x}$ and $Y\mid A{=}0, \boldsymbol{X}{=}\boldsymbol{x}$. Then, one can, for example, aim to derive the TBR that equals $\mathbb{P}(Y^1-Y^0 \leq 0 \mid \boldsymbol{X}{=}\boldsymbol{x} )$ when lowering the outcome is beneficial,  or the THR $\mathbb{P}(Y^1-Y^0 > 0 \mid \boldsymbol{X}{=}\boldsymbol{x})$. Moreover, by marginalizing over $\boldsymbol{X}$, one can study the proportion of individuals with particular effect sizes. For example, the proportion of individuals that will experience at least twice the average effect $\mathbb{P}(Y^1-Y^0 > 2 \cdot \mathbb{E}[Y^1-Y^0])$ as we will do in the case study in Section \ref{CH3sec6}. 

\section{Causal mixed models}\label{CH3sec5}
We continue by focusing on fitting the unknown $F_{Y \mid A{=}1, \boldsymbol{X}{=}\boldsymbol{x}}$, and $F_{Y \mid A{=}0, \boldsymbol{X}{=}\boldsymbol{x}}$ from data. Under assumptions \ref{CH2A2}, \ref{CH2A3} and \ref{CH2A1}, $Y \mid A{=}a, \boldsymbol{X}{=}\boldsymbol{x}$ is equal in distribution to $Y^{a} \mid \boldsymbol{X}{=}\boldsymbol{x}$ thus by SCM \eqref{CH3SCMsurv2}, 
\begin{equation}\label{CH3eq:truth}
Y_{i} = f_{Y0}(\boldsymbol{X}_{i}) + N_{Yxi} + (f_{Y1}(\boldsymbol{X}_{i}) + U_{1xi})A_{i},\text{ } N_{Yxi} \sim F_{Yx}(\boldsymbol{X}_{i}) \text{ and } U_{1xi} \sim F_{U_{1x}}(\boldsymbol{X}_{i}), 
\end{equation} where $f_{Y0}(\boldsymbol{x}) = \theta + \mathbb{E}[N_{Y}\mid \boldsymbol{X}{=}\boldsymbol{x}]$, $f_{Y1}(\boldsymbol{x}) = \tau + \mathbb{E}[U_{1}\mid \boldsymbol{X}{=}\boldsymbol{x}]$ 
and the distributions $F_{N_{Yx}}(\boldsymbol{x})$ and $F_{U_{1x}}(\boldsymbol{x})$ can depend on $\boldsymbol{x}$ but have zero mean for all $\boldsymbol{x}$. If $N_{Yx}$ does not depend on $\boldsymbol{X}$, then $f_{Y0}(\boldsymbol{X})$ is the prognostic score of $Y^{0}$ since $Y^{0} \independent \boldsymbol{X} \mid f_{Y0}(\boldsymbol{X})$ \citep{Hansen2008}. If also $U_{1x}$ is independent of $\boldsymbol{X}$, then $\{f_{Y0}(\boldsymbol{X}), f_{Y1}(\boldsymbol{X})\}$ is the prognostic score of $Y^{1}$. 
  
Note that $\int f_{Y1}(\boldsymbol{x}) dF_{\boldsymbol{X}}(\boldsymbol{x})$ equals the ATE $\tau$,  $f_{Y1}(\boldsymbol{x})$ equals the CATE for $\boldsymbol{X}{=}\boldsymbol{x}$, and the ICE for individual $i$ equals $f_{Y1}(\boldsymbol{X}_{i})+U_{1xi}$.

The distribution of $(U_{1x}, N_{Yx})$ in \eqref{CH3eq:truth} and the functional forms of $f_{Y0}$ and $f_{Y1}$ are unknown. One can try to fit the observed data using a (flexible) mixed model with a random exposure effect $Z_{1}$,
\begin{equation}\label{CH3mixedmodel}
Y_{i}  =  \beta_{0}(\boldsymbol{X}_{i}) + \varepsilon_{i} + (\beta_{1}(\boldsymbol{X}_{i}) + Z_{1i}) A_{i}, \text{ } \varepsilon_{i} \sim F_{\varepsilon}(\boldsymbol{X}_{i}) \text{ and } Z_{1i} \sim F_{Z_{1}}(\boldsymbol{X}_{i}), 
\end{equation} 

where the distributions $F_{\varepsilon}(\boldsymbol{x})$ and $F_{Z_{1}}(\boldsymbol{x})$ can depend on $\boldsymbol{x}$ but have zero mean for all $\boldsymbol{x}$, e.g., heteroskedastic residuals, but we assume $Z_{1i} \independent \varepsilon_{i} \mid \boldsymbol{X}_{i}$. At this point, the latter independence is thus the only restriction of the associational model \eqref{CH3mixedmodel}. Note that the functions $f_{Y0}$, $f_{Y1}$,  and the distributions of $U_{1x}$ and $N_{Yx}$ follow from SCM \eqref{CH3SCMsurv2}, while the functions $\beta_{0}$, $\beta_{1}$, and the distributions of $Z_{1}$ and $\varepsilon$ are part of the statistical model that is fitted to the observed data. If assumptions \ref{CH2A2}, \ref{CH2A3}, \ref{CH2A1} and \ref{CH2A4} apply, and the model \eqref{CH3mixedmodel} is well specified\footnote{Well specified in the sense that $\beta_{0}$ and $\beta_{1}$ have the same functional form as $f_{Y0}$ and $f_{Y1}$ respectively, while $\epsilon$ and $\epsilon+Z_{1}$ follow the same conditional distribution (on $\boldsymbol{X}$) as $N_{Yx}$ and $N_{Yx}+U_{1x}$ respectively.}, then the distribution of  $\beta_{1}(\boldsymbol{X}){+}Z_{1}{\mid}\boldsymbol{X}{=}\boldsymbol{x}$ equals the distribution of $Y^{1}-Y^{0}{\mid} \boldsymbol{X}{=}\boldsymbol{x}$ by Proposition \ref{Cor1}. There is a lot of research on non-parametric (machine learning) methods to estimate the functionals $\beta_{0}$ and $\beta_{1}$ \cite{Post2024}. In the remainder of the paper, we will focus on estimating the (conditional) distributions of $\epsilon$ and $Z_{1}$. 

\subsection{Methods to fit the random effects model}\label{sec:methods}
Linear mixed models (LMMs) are often used in epidemiological studies and particularly the fitting of LMMs with (independent) Gaussian random effects ($Z_{1}$) and Gaussian residual $(\varepsilon)$ is implemented in standard software packages, e.g.~\texttt{PROC MIXED} in \texttt{SAS} and the \texttt{lmer} package in \texttt{R}. Letting the variance components of $\varepsilon$ and $Z_{1}$ depend on $\boldsymbol{X}$ is possible. The Gaussian model might be fitted using likelihood-based or Bayesian methods \citep{Browne2006}. Note that when there exists $\boldsymbol{x}$ such that $U_{1}{\mid} \boldsymbol{X}{=}\boldsymbol{x}$ or $N_{Y} {\mid} \boldsymbol{X}{=}\boldsymbol{x}$ are not Gaussian distributed, the Gaussian mixed model is misspecified. Then, the $Z_{1}$ distribution would be a biased estimate of the conditional ICE distribution. However, even in that case, a Gaussian random effects fit can be helpful to study whether there is any remaining variability in ICE within a subpopulation where $\boldsymbol{X}{=}\boldsymbol{x}$. When the variance component of the random exposure effect is small, and assumptions \ref{CH2A2}, \ref{CH2A3}, \ref{CH2A1}, and \ref{CH2A4} apply, one could use the CATE as an appropriate proxy for the ICE. 
If the size of the variance component is considerable, the TBR and THR in the subpopulation could be estimated from the LMM fit, as proposed by \citet{Yin2018}. These TBR and THR estimates will only be accurate when $U_{1}{\mid} \boldsymbol{X}$ and $N_{Y} {\mid} \boldsymbol{X}$ are (approximately) Gaussian distributed. In the case of (only) misspecified random-effects distributions in generalized linear mixed models (GLMMs), the fixed effect parameters of the marginalized population characteristics are still unbiased, but the standard errors of the estimates are affected \citep{McCulloch2011}.

When interested in the ICE distribution, the distribution of the random effect should be well-specified and thus not restricted to be Gaussian. \citet{Verbeke1996} suggested modeling the random effect as a mixture of normals with an unknown number of components. This model can be fitted using an Expectation-Maximization algorithm \citep{Verbeke1996}, and an alternative estimation procedure has been proposed by  \citet{Proust2005}. To study features of between-individual variation, \citet{Zhang2001} proposed to fit an LMM by approximating the density of the random effect by a semi-nonparametric representation. For both approaches, an optimal tuning parameter is often selected based on information criteria \citep{Zhang2001, Proust2005}. 

The Gaussian mixture distribution for the random effect can also be fitted in a Bayesian Framework \citep{Kleinman1998}. As the ICE distribution is unknown, information criteria can be used to set the optimal number of components \citep[Chapter 22]{Gelman2021}. A one-step approach using a uniform prior distribution on the number of components of the mixture distribution of the random effect has been proposed to avoid model selection \citep{Ho2008}. Model selection is also not necessary when one does fix the number of components to a `large' constant $K$ while using a symmetric Dirichlet prior, with an appropriate parameter $\boldsymbol{\alpha}$, for the $K$-class probabilities and rely on the natural penalization induced by Bayesian approaches \citep[Chapter 22]{Gelman2021}. When $K$ is larger than the actual number of components, the mixture parameters are not identifiable. However, this is not an issue when the distribution is the main object of interest, and the parameters of the mixture components are not of interest \citep{Rousseau2011}. Moreover, simulation studies suggest that $\boldsymbol{\alpha}=(K^{-1}, \hdots, K^{-1})$ will suffice to empty components that are not necessary to fit a Gaussian mixture with an unknown number of components \citep[Chapter 22]{Gelman2021}. Indeed, for univariate random effects, under regularity assumptions, it has been shown that for $\alpha<1$, the posterior distribution will have empty components when $K$ is larger than the actual number of components, while this is not the case for $\alpha>1$ \citep{Rousseau2011}.  

Finally, the random effect distribution can be modeled non-parametrically. For example, a Bayesian nonparametric fit of a hierarchical model can be obtained using a Dirichlet process prior \citep{Dunson2009} or a truncated Dirichlet process prior \citep{Ohlssen2007} respectively, for the distribution of the random effect. Despite the existence of all methods, precise estimation of the distribution of a latent variable remains challenging and is sensitive to the proposed model. 

\subsection{Example ICE distribution estimation}\label{CH3sec:bayes}
\begin{figure}[t]
	\centering
	\begin{subfigure}{.325\textwidth}
		\resizebox{1\linewidth}{!}{\includegraphics{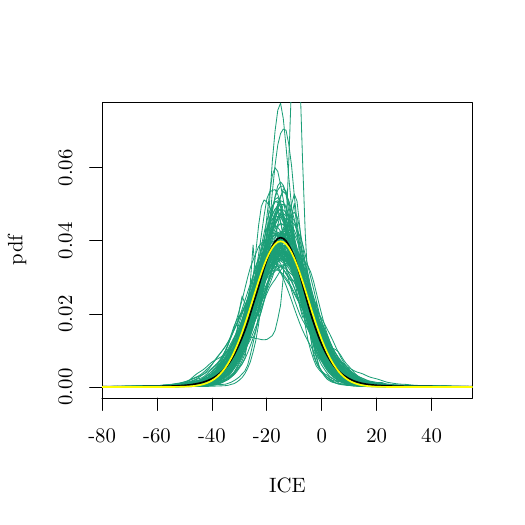}}
		\caption{}\label{CH33d}	
	\end{subfigure} 
	\begin{subfigure}{.325\textwidth}
		\resizebox{1\linewidth}{!}{\includegraphics{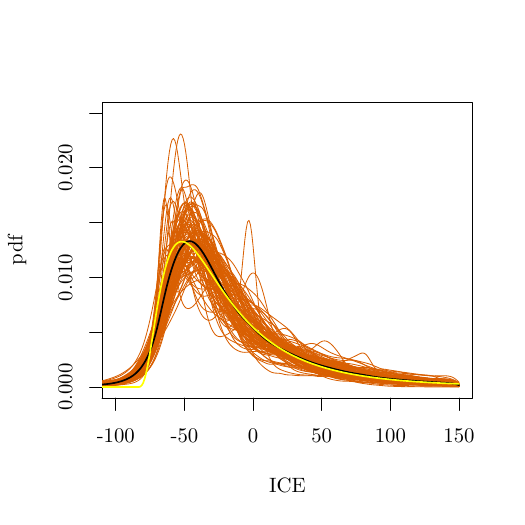}}
		\caption{}\label{CH33e}	
	\end{subfigure}
	\begin{subfigure}{.325\textwidth}
		\resizebox{1\linewidth}{!}{\includegraphics{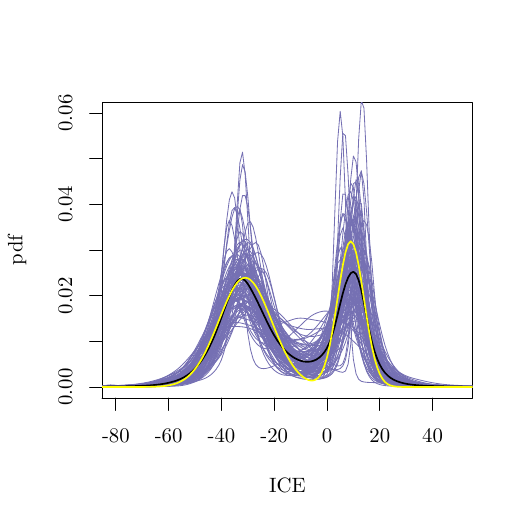}}
		\caption{}\label{CH33f}	
	\end{subfigure}
	\caption{Estimated $Z_{1}$ distributions for $100$ simulations for underlying ICE distributions $\tau+U_{1x}$, where $\tau=-15$, $U_{1x}~{\sim}~\mathcal{N}(0,10^2)$ (left), or  $U_{1x} ~{\sim}~ \text{LN}(4,\sqrt{0.5}^{2})-\exp\left(4+0.5 \sqrt{0.5}^{2}\right)$ 
 (middle) or a two-component Gaussian mixture such that with probability $p=0.6$ $U_{1}~{\sim}~ \mathcal{N}(-16,10^{2})$ and otherwise $U_{1}~{\sim}~ \mathcal{N}(24,5^{2})$ (right). The actual density for $\tau+U_{1}$(yellow) and the mean of the estimated $Z_{1}$ densities (black) are presented. Data for individuals were simulated using Equation \eqref{CH3eq:truth}, where $N_{Yx} \sim \mathcal{N}(0, 50)$, $f_{Y0}(\boldsymbol{x})=  \theta_{0}+ x\theta_{1}$,  $f_{Y1}(\boldsymbol{x})=\tau$, where $\theta_{0}=120$ and the effect of the confounder was $\theta_{1}=5$. The confounder $X$ had the value $-0.3$ with probability $0.7$ and value $0.7$ with probability $0.3$ and thus a zero mean. The probability of exposure equaled $\text{logit}(-3+3 X_{i})$ for individual $i$. Finally, $N_{Yx}, U_{1x} \independent A$ and  $U_{1x} \independent N_{Yx}$.} \label{CH3FIG3}
\end{figure}

To demonstrate the use of a causal mixed model, we will next consider a simple example with data simulated for $1000$ individuals using Equation \eqref{CH3eq:truth} with $f_{Y0}(\boldsymbol{x})=  \theta_{0}+ x\theta_{1}$, while  $N_{Yx}, U_{1x} \independent A$,  $U_{1x} \independent N_{Yx}$, $N_{Yx}$ is Gaussian distributed, and $U_{1x}$ follows a Gaussian distribution, a log-normal distribution or a mixture of two Gaussians respectively. For simplicity, we will assume that only one binary confounder exists that is no modifier (on the additive scale, $f_{Y1}(\boldsymbol{x})=\tau$). Details on the parameters used can be found in the caption of Figure \ref{CH3FIG3}. 
 
We will model the ICE distribution with a Gaussian mixture that we fit using the Bayesian method with an upper bound for the number of components (discussed in Section \ref{sec:methods}) since this one-step approach can be easily implemented in both \texttt{SAS} and \texttt{R}. Moreover, since we are interested in estimating the entire distribution of the ICE, a Bayesian method has the advantage that (pointwise) uncertainty quantification of the ICE density can be directly obtained. On the contrary, when using a frequentist approach, the uncertainty of the density should be derived from the uncertainty in the model parameters. 

We model the observed data as
\begin{equation}\label{CH3LVmodel}
Y_{i} = \beta_{0} + X_{i}\beta_{X} + A_{i}Z_{1i}+\varepsilon_{i},
\end{equation} where $Z_{1i}$ follows a $K$-dimensional Gaussian mixture distribution with mixing probabilities $\boldsymbol{p}$, component means $\boldsymbol{\mu}$ and component variances $\boldsymbol{\tau^{2}}$ (i.e., $Z_{1i}~{\sim}~ \text{GM}(\boldsymbol{p}, \boldsymbol{\mu}, \boldsymbol{\tau}^{2})$), $\varepsilon_{i} ~{\sim}~ \mathcal{N}(0,\sigma^{2})$, $Z_{1i} \independent \varepsilon_{i}$ and $Z_{1i}, \varepsilon_{i} \independent X_{i}$. For $1\leq j \leq K$, the recommended non-informative priors \citep[Chapter 22]{Gelman2006} 
\begin{align}
\mu_{j}, \beta_{0}, \beta_{X} & ~{\sim}~ \mathcal{N}(0, 10^5) \\
\tau_{j}, \sigma & ~{\sim}~ \text{Uni}[0,100],
\end{align} and the weakly informative prior
\begin{equation}
\boldsymbol{p} ~{\sim}~ \text{Dir}(\alpha, \hdots, \alpha),
\end{equation} are used.
When \eqref{CH3LVmodel} is well-specified and assumptions \ref{CH2A2}, \ref{CH2A3}, \ref{CH2A1} and \ref{CH2A4} apply, by Proposition \ref{Cor1}, $\mathbb{P}(Y^1 - Y^0 \leq y)$ equals $\mathbb{P}(Z_{1} \leq y)$. Since $Z_{1}$ is a univariate random variable, $K=5$ should suffice to capture important characteristics, i.e., skewness and multi-modality, of the ICE distribution as suggested elsewhere \citep[Section 22.4]{Gelman2021}. As we do not have prior knowledge about the ICE distribution, we should be careful not to select a too small $\alpha$ that would favor distributions closer to a normal distribution via the weakly informative prior. The prior distribution of the number of prominent components, i.e.,  $p_{j} \in \boldsymbol{p}{:}~ p_{j}>0.10$, for different levels of $\alpha$ is presented in Figure \ref{CH3Dprior}. Here, we use $\alpha{=}0.5$, giving rise to Jeffreys prior for the labeling distribution \citep{Rousseau2011}. In this work, we use \texttt{PROC MCMC} in \texttt{SAS} to perform the Bayesian analysis. 
During every Monte-Carlo iteration, for each individual, a latent modifier is sampled from the Gaussian mixture based on the parameters sampled for that specific iteration. Finally, the posterior $Z_{1}$ distribution is presented using a kernel density estimation of all sampled $Z_{1}$. The posterior distributions of $Z_{1}$ are presented for $100$ simulated datasets and different underlying ICE distributions in Figure \ref{CH3FIG3}.  \noindent In expectation (black lines), the characteristics of the ICE distribution are well recovered. For illustration, for an arbitrary dataset, the kernel density estimates of the sampled $Z_{1}$ per Markov chain Monte Carlo (MCMC) iteration (grey) are presented together with the posterior distributions of $Z_{1}$ (colored) and the underlying $Y^{1}-Y^{0}$ distribution (yellow) as Figures \ref{CH33a} to \ref{CH33c} in the Appendix. 

The mean values of the posterior means and the coverage of the Bayesian credible sets are presented in Table \ref{CH3tab1} for $\mathbb{P}(Z_{1}>0)$ and the quantiles of the distribution of $Z_{1}$. For these datasets with $1000$ individuals, the averages of the posterior means for all three ICE distributions are close to the actual levels of $\mathbb{P}(Y^1-Y^0>0)$ and the quantiles of the ICE distribution considered. Note that $\mathbb{P}(Y^1-Y^0>0)$ is equal to the TBR if higher outcomes are beneficial and equal to the THR otherwise. Furthermore, the coverages of the Bayesian credible sets for this sample size are pretty close to $0.95$, although for the non-Gaussian $U_{1}$ distributions the coverage tends to be slightly lower.

\begin{table}[t]
	\centering
	\caption{ Actual values of $\mathbb{P}(Y^{1}-Y^{0}>0)$ and the $5\%$, $25\%$, $50\%$, $75\%$ and $95\%$ quantiles for the different ICE distributions. For these characteristics, the average of the posterior means, each estimated on $1000$ individuals, and the coverage of the $95\%$ Bayesian credible sets, based on the $100$ simulations, are presented.}\label{CH3tab1}
	
	\begin{tabular}{c||ccc|ccc|ccc}
		& \multicolumn{3}{c|}{Actual value} & \multicolumn{3}{c|}{Average posterior means} & \multicolumn{3}{c}{Coverage credible sets} \\
		& (a)    & (b)     & (c)    & (a)       & $\text{(b)}^{*}$        & (c)         & (a)          & $\text{(b)}^{*}$             & (c)          \\ \hline
		$\mathbb{P}(Y^1-Y^0>0)$          & 0.07   & 0.27    & 0.39   &   0.08        & 0.27       & 0.36        &    0.95          & 	0.94         & 0.90        \\
		$\alpha_{0.05}$          & -31.44   & -68.04 &-44.83   & -33.48   & -71.96 &-45.58      & 0.90  & 0.94   & 0.90  \\
		$\alpha_{0.25}$          & -21.74 & -51.21  &  -33.10      &  -21.72         & -50.19       & -33.13       & 0.92             & 	 0.89        & 0.91  \\
		$\alpha_{0.5}$          &  -15 & -30.51 &  -21.33  &     -14.88   & -30.31       & -18.71       & 0.95            & 	0.89        &  0.90 \\
		$\alpha_{0.75}$          &  -8.25 & 2.86 & 7.41       &   -8.14       & 4.14       &  6.83      & 0.96            & 	0.93         &   0.88      \\
		$\alpha_{0.95}$          &  1.44  & 89.60 & 14.75      &    3.42       & 90.50       & 15.90       & 0.97            & 	0.89         & 0.86 	\\
	\end{tabular} 
\end{table}

\begin{figure}[t]
	\centering
	\begin{subfigure}{.4\textwidth}
		\resizebox{1\linewidth}{!}{\includegraphics{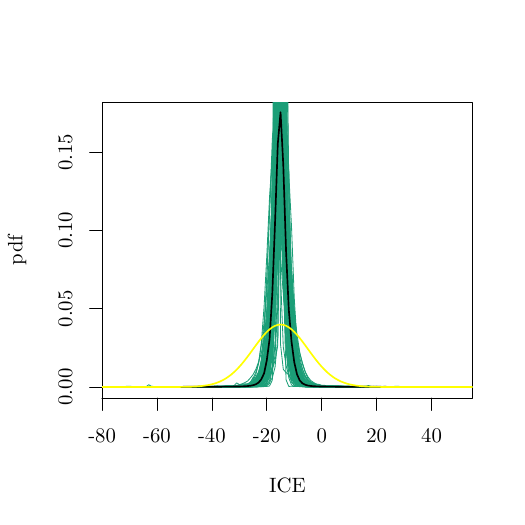}}
		\caption{}\label{CH3FIGsensA}
	\end{subfigure} 
	\begin{subfigure}{.4\textwidth}
		\resizebox{1\linewidth}{!}{\includegraphics{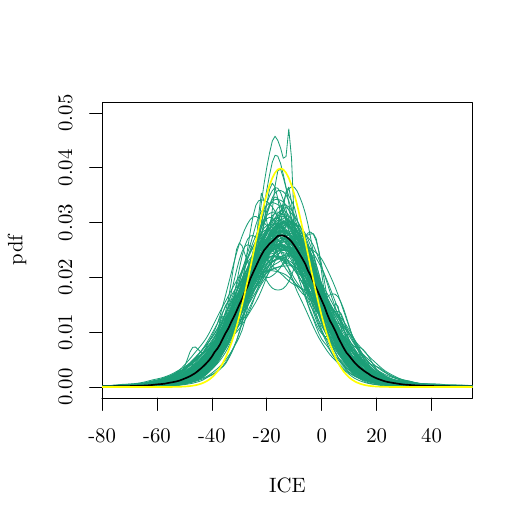}}
		\caption{}\label{CH3FIGsensB}	
	\end{subfigure}
	\caption{Estimated $Z_{1}$ distributions for $100$ simulations for underlying ICE distributions $\tau+U_{1x}$, where $\tau=-15$ and $(U_{1x}, N_{Yx})$ is bivariate normal distributed with zero means, variance $10^2$ and $50$ respectively and correlation $\rho=-0.75$ (a) or $\rho=0.75$ (b). The actual density for $\tau+U_{1x}$ (yellow) and the mean of the estimated $Z_{1}$ densities (black) are presented. Data for individuals were simulated using Equation \eqref{CH3eq:truth}, where $N_{Yx} \sim \mathcal{N}(0, 50)$, $f_{Y0}(\boldsymbol{x})=  \theta_{0}+ x\theta_{1}$,  $f_{Y1}(\boldsymbol{x})=\tau$, where $\theta_{0}=120$ and the effect of the confounder was $\theta_{1}=5$. The confounder $X$ had the value $-0.3$ with probability $0.7$ and value $0.7$ with probability $0.3$ and thus a zero mean. The  probability of exposure equaled $\text{logit}(-3+3 X_{i})$ for individual $i$ and $N_{Yx}, U_{1x} \independent A$.} \label{CH3FIGsens}
\end{figure}

In practice, the model itself should be validated after verifying the proper convergence of the Markov chains. The causal assumptions \ref{CH2A2}, \ref{CH2A1} and \ref{CH2A4} cannot be verified, but 
the associational model can be validated using the observational data, as we will elaborate on in Section \ref{CH3sec6}. It is essential to realize that when Assumption \ref{CH2A4} is violated, the ICE distribution is not identified by the $Z_{1}$ distribution. If $Y^{0}$ and $Y^{1}-Y^{0}$ are negatively correlated, the variability in the ICE distribution will be underestimated by the variability in the $Z_{1}$ distribution. In Figure \ref{CH3FIGsensA}, this is illustrated for data simulated from the system discussed in this section, but now the correlation of $N_{Yx}$ and $U_{1x}$ is $\rho=-0.75$. Similarly, for $\rho=0.75$, $Y^{0}$ and $Y^{1}-Y^{0}$ are positively correlated, and the variability of the ICE distribution is overestimated, see Figure \ref{CH3FIGsensB}.


\subsection{Confounding-effect heterogeneity}\label{CH3confhet}
As stated in Equation \eqref{CH3eq:truth}, the distribution of $N_{Yx}$ and $U_{1x}$ can depend on $\boldsymbol{X}$. Then, also $\varepsilon$ and $Z_1$ should depend on $\boldsymbol{X}$ to have a well-specified model \eqref{CH3mixedmodel}. This is particularly important when the distributions of $N_{Yx}$ and $U_{1x}$ depend on confounders since the fitted distribution of $Z_1$ will represent the difference in the shape of the conditional distribution between exposed and unexposed individuals. Thus, when we do not appropriately adjust for confounders, $Z_1$ will never represent the $U_{1x}$ distribution. 

\begin{figure}[t]
	\centering
	\begin{subfigure}{.4\textwidth}
		\resizebox{1\linewidth}{!}{\includegraphics{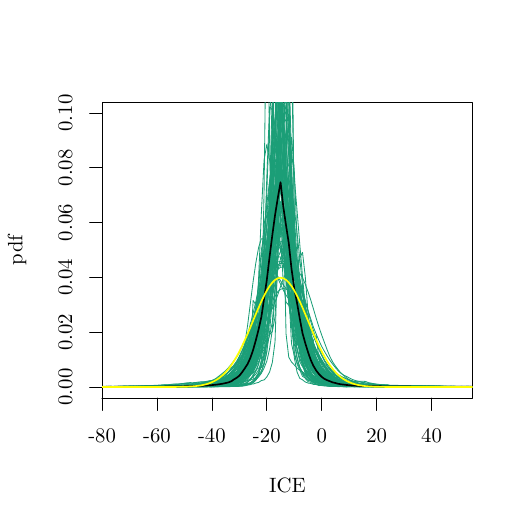}}
		\caption{}\label{CH3FIGchA}
	\end{subfigure} 
	\begin{subfigure}{.4\textwidth}
		\resizebox{1\linewidth}{!}{\includegraphics{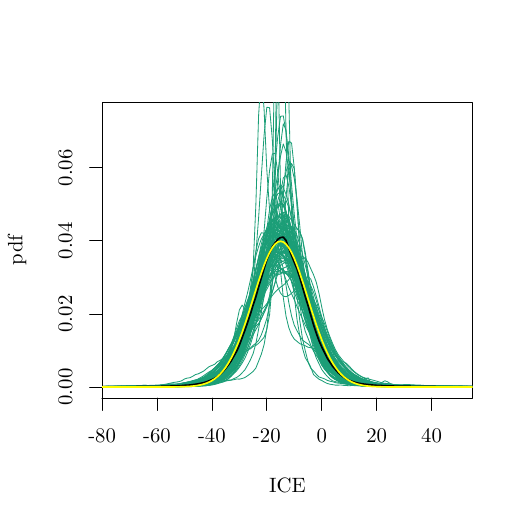}}
    \caption{}\label{CH3FIGschB}	
	\end{subfigure}
	\caption{Estimated $Z_{1}$ distributions after fitting model \eqref{CH3LVmodel} (a) or \eqref{CH3LVmodelfix} (b) for $100$ simulations for underlying ICE distributions $\tau+U_{1x}$, where $\tau=-15$ and $U_{1x}~{\sim}~\mathcal{N}(0,10^2)$. The actual density for $\tau+U_{1}$(yellow) and the mean of the estimated $Z_{1}$ densities (black) are presented. Data for individuals were simulated using Equation \eqref{CH3eq:truth}, where $N_{Yx} $ Gaussian distributed, $f_{Y0}(\boldsymbol{x})=  \theta_{0}+ x\theta_{1}$,  $f_{Y1}(\boldsymbol{x})=\tau$, where $\theta_{0}=120$ and the effect of the confounder was $\theta_{1}=5$. The confounder $X$ had the value $-0.3$ with probability $0.7$ and value $0.7$ with probability $0.3$ and thus a zero mean. The variance of $N_{Yx}$ was equal to $50$ for $X=0.7$ and $50 + 10^2$ for $X=-0.3$ and the probability of exposure equaled $\text{logit}(-3+3 X_{i})$ for individual $i$ and $N_{Yx}, U_{1x} \independent A$.} \label{CH3FIGch}
\end{figure}

For causal inference from observational data, it is often necessary to adjust for features so that Assumption \ref{CH2A1} applies. If the interest is in the ATE, we in principle only need to take into account the confounder's effect on the outcome's mean, i.e., we need to adjust for $\boldsymbol{X}$ so that 
$$\mathbb{E}[Y^a \mid \boldsymbol{X}=\boldsymbol{x}] = \mathbb{E}[Y \mid \boldsymbol{X}=\boldsymbol{x}, A=a].$$ However, if one is interested in the ICE distribution, then it is necessary to consider the effect of the confounder on the entire distribution of the outcome, i.e., adjust for $\boldsymbol{X}$ so that Assumption \ref{CH2A1} applies and
$$F_{Y^a \mid \boldsymbol{X}=\boldsymbol{x}}(y) = F_{Y \mid \boldsymbol{X}=\boldsymbol{x}, A=a}(y).  $$
Otherwise, a difference in (the shape of the) distribution between the exposed and unexposed individuals can be caused by the non-exchangeable confounders. Then, the estimated distribution of $Z_{1}$ in \eqref{CH3mixedmodel} is a mixture of the heterogeneity in the effects of the confounders and the exposure effect. By verifying whether the proposed distribution fits well in sub-samples grouped by the levels of the confounders, one can decide whether accounting for the effect on the mean is sufficient. 
The effects of a confounder on the outcome of different individuals might follow a non-degenerate distribution. If so, then confounding-effect heterogeneity exists. An (extreme) example of confounding-effect heterogeneity is the case where only the outcome of a proportion of individuals is affected by the confounder \citep{Bonvini2020}. For another example, let's again consider the example presented in Section \ref{CH3sec:bayes}, but now we let $N_{Yx}$ depend on $X$ so that $N_{Yx} \mid X=0.7$ and $N_{Yx} \mid X=-0.3$ are Gaussian distributed with variance $50$ and $50 + 10^2$ respectively. Then, model \eqref{CH3LVmodel} is misspecified, and as a result, the variability of the ICE distribution is underestimated, as presented in Figure \ref{CH3FIGchA}. The variability is underestimated since among the unexposed $X=-0.3$ is more prevalent than among the exposed. 

In the case of confounding-effect heterogeneity, the distribution of $\varepsilon$ in \eqref{CH3mixedmodel} should depend on the level of the confounders. Indeed, when model \eqref{CH3LVmodel} is extended as
\begin{equation}\label{CH3LVmodelfix}
Y_{i} = \beta_{0} + X_{i}\beta_{X} + \tilde{X}_{i}Z_{Xi} + A_{i}Z_{1i}+\varepsilon_{i},
\end{equation} where $\tilde{X}_{i}$ equals $0$ if $X_{i}=0.7$ and $1$ otherwise and $Z_{Xi} ~{\sim}~ \mathcal{N}(0,\tau_{X}^{2})$, the model is well specified and the ICE distribution can be accurately represented with the $Z_{1}$ distribution as presented in Figure \ref{CH3FIGschB}.  

Similarly, the $Z_{1}$ distribution does not represent the ICE distribution when the $N_{Yx}$ distribution cannot be approximated with the proposed distribution of $\varepsilon$. Then, the $Z_{1}$ distribution obtained by fitting model \eqref{CH3mixedmodel} is a mixture of $U_{1x}$ and the deviation of the distribution of $N_{Yx}$ from the proposed distribution for $\varepsilon$, such that the model will better fit the data of the exposed individuals. This problem can be verified by evaluating the fit to the data of the unexposed individuals and might be solved using a more flexible distribution for $\varepsilon$.

\newpage
\section{Case study: the Framingham Heart Study}\label{CH3sec6}
In this section, we consider heterogeneity in the effect of non-alcoholic fatty liver disease on cardiac structure and function, as studied by \citet{Chiu2020} in the Framingham Heart Study (FHS) population. We will work with a subset of the FHS third generation and offspring cohorts $(n=2352)$. To illustrate our proposed method, we focus on the effect of Hepatic Steatosis (fatty liver disease) on the left ventricular filling pressure (LVFP), a clinical precursor to heart failure. Hepatic Steatosis was defined as having a liver phantom ratio (measured fat attenuation in three areas of the liver divided by a phantom calibration control) exceeding $0.33$. It was estimated to increase the expected LVFP by $0.46$, with a $95 \%$ confidence interval equal to ($0.26, 0.65$), after adjustment for confounding \citep{Chiu2020}. 

The sample standard deviation of the LVFP equals $2.27$ for individuals exposed to Hepatic Steatosis and only $1.74$ for those who were not. This difference might be the result of causal effect heterogeneity. As the Bayesian analysis discussed in Section \ref{CH3sec:bayes} is computationally intensive, we started by fitting a traditional Gaussian LMM to investigate which candidate confounders affected the estimated mean or variance of the effect of Hepatic Steatosis on LVFP. Confounders accounted for in the original study were age, sex, smoking, alcohol use, diabetes, systolic blood pressure (SBP), use of antihypertensive medication (HRX), use of lipid-lowering medication, total cholesterol, high-density lipoprotein cholesterol, triglycerides, and fasting glucose. The relevant confounders were age, sex, diabetes, SBP, and HRX (details can be found in Section \ref{CH3CH3:A0} of the Appendix). 

In this case study, our focus is on the population ICE distribution; thus, we do not consider modifiers that are no confounders. We have fitted the model 
\begin{equation}\label{CH3LVmodel3}
Y_{i} = \beta_{0} + \boldsymbol{X}_{i}\boldsymbol{\beta}_{X}^T  + \left(\boldsymbol{X}_{i}\boldsymbol{\beta_{\text{XA}}}^T + Z_{1i}\right)A_{i}+\varepsilon_{i}, 
\end{equation} where $Z_{1i} ~{\sim}~ \text{GM}(\boldsymbol{p}, \boldsymbol{\mu}, \boldsymbol{\tau}^{2})$, with $K=5$, $\varepsilon_{i} ~{\sim}~ \text{GM}(\tilde{\boldsymbol{p}}, \boldsymbol{\tilde{\mu}}, \boldsymbol{\tilde{\tau}}^{2})$ with $\tilde{{}K}=3$ while restricting $\mathbb{E}[\varepsilon_{i}]=0$, and $Z_{i} \independent \varepsilon_{i}$. We believe that it was appropriate to model the residual (and thus $Y{\mid} A{=}0, \boldsymbol{X}{=}\boldsymbol{x}$) with a more flexible distribution as we do not condition on any features other than adjusting for confounders such that the distribution is expected to be more complex than Gaussian. This is important since misspecification of the residual distribution will affect the estimated distribution of $Z_{1}$ as explained at the end of Section \ref{CH3confhet}. Alternative flexible error distributions have been presented in the literature \citep{Ghidey2010, Shahn2017, Rubio2018}.

The model was fitted using the Bayesian method discussed in Section \ref{CH3sec:bayes}. The initial values for the parameters in the Markov chain were based on the final Gaussian LMM used to select the confounders and can be found in Table \ref{CH3tab:A0e} in the Appendix. Per chain, we have used $100{,}000$ burn-in iterations, followed by $500{,}000$ MCMC iterations. We have used a thinning rate of $100$ to save computer memory space. In total, $5$ chains, 
each contained $5000$ (thinned) MCMC iterations that were saved. Each chain was split into two pieces, and we investigated the convergence of the $\boldsymbol{X}_{i}\boldsymbol{\beta_{\text{XA}}}^T + Z_{1i}$ distribution for the first unexposed individual in the dataset. Convergence of the Markov chains was supported by an estimated R-hat convergence diagnostic, as defined by \citet{Vehtari2021}, equal to $\hat{R}{=}1.01$ and the estimated bulk and tail effective sample sizes equal to $2.5 {\cdot} 10^4$ and $1.8 {\cdot} 10^4$ respectively. The trace plot for the $\boldsymbol{X}_{i}\boldsymbol{\beta_{\text{XA}}}^T + Z_{1i}$ of this (and another) unexposed individual can be found in Figure \ref{CH3fig:traceplot} in the Appendix. 



\begin{figure}[t!]
	\centering
	\includegraphics[width=\linewidth]{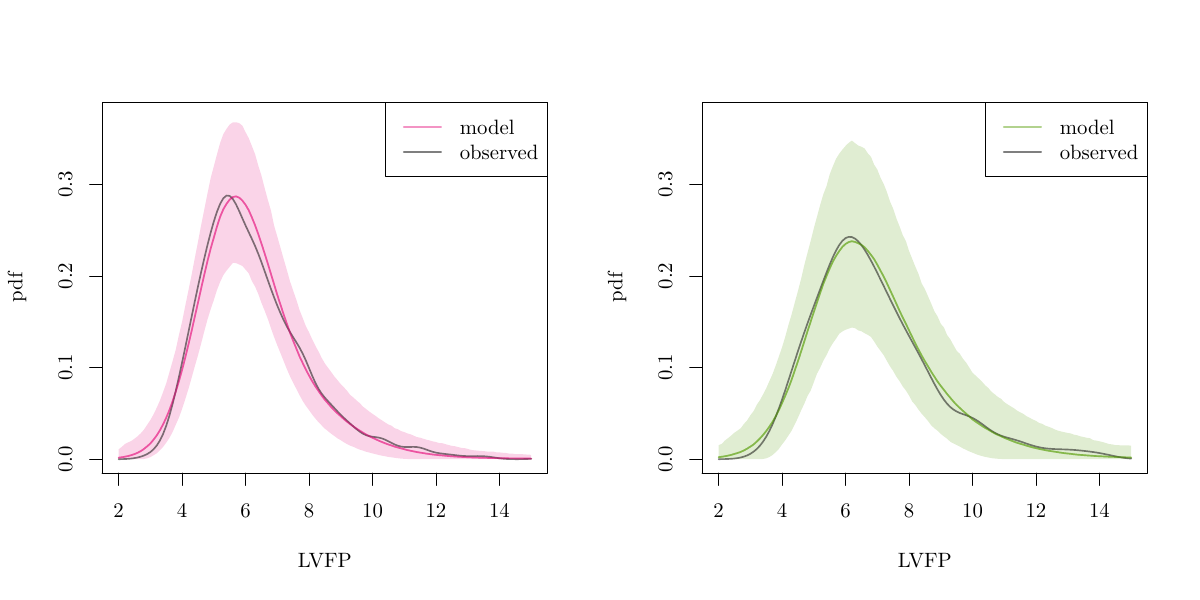}
	\caption{Posterior predictive distribution of the LVFP for the model   \eqref{CH3LVmodel3}, with corresponding $95 \%$ Bayesian credible intervals and the kernel density of the observed data for unexposed (left) and exposed individuals (right). }
	\label{CH3fig:fattylivernoc}
\end{figure}

We should ensure that the model accurately fits the observed data to draw inferences on the ICE distribution. To validate the model, we study the posterior predictive distribution of the LVFP for both the exposed and unexposed individuals ($Y{\mid} A{=}a)$ as presented in Figure \ref{CH3fig:fattylivernoc}.  The distribution of the outcomes of the unexposed individuals ($Y{\mid} A{=}0$) is well-fitted using a 3-component Gaussian mixture residual. Furthermore, a 5-component Gaussian mixture for the individual effect distribution is flexible enough to fit well the observations of the exposed individuals ($Y {\mid} A{=}1$). 

As shown in Table \ref{CH3varcas}, the standard deviation of the LVFP is higher for individuals who are either older, female, have diabetes, have higher blood pressure, or use antihypertensive medication for both exposure groups.  
\begin{table}[t]
	\begin{tabular}{cc|cc|cc|cc|cc|cc}
		&   & \multicolumn{2}{c|}{Age\textgreater{}49} & \multicolumn{2}{c|}{Sex} & \multicolumn{2}{c|}{Diabetes} & \multicolumn{2}{c|}{SBP\textgreater{}120} & \multicolumn{2}{c}{HRX} \\
		&   & 0                  & 1                  & 0          & 1          & 0             & 1            & 0                   & 1                  & 0                    & 1\\ 
		\midrule
		\multirow{2}{*}{Hepatic Steatosis} & 0 & 1.4           & 1.9          & 1.5   & 1.9   & 1.7      & 2.0     & 1.5           & 1.9          & 1.6            & 2.1            \\
		& 1 & 1.7           & 2.6           & 1.8   & 2.6   & 2.2      & 2.5     & 2.0          & 2.3           & 1.9             & 2.7      \\     
	\end{tabular}
 \centering
	\caption{Sample standard deviation of the LVFP in sub-samples partitioned by the (dichotomized) confounders and the exposure. }\label{CH3varcas}
\end{table} In Section \ref{CH3confhet}, we have explained how confounding-effect heterogeneity can affect the distribution of $Z_{1}$. The validation of the fit of the $Y {\mid} \boldsymbol{X}{=}\boldsymbol{x}$ distribution for all dichotomized levels of the confounders ($>$ median value) is thus a crucial part of the analysis. The posterior predictive distributions in all groups of dichotomized confounders are presented in Section \ref{CH3appNOC} of the Appendix. 
Overall, we conclude that significant confounding-effect heterogeneity seems to be absent. In Section \ref{CH3sensa}, we demonstrate how a model may be fitted when confounding-effect heterogeneity is present.   

\begin{figure}[t]
	\centering
	\includegraphics[width=0.75\linewidth]{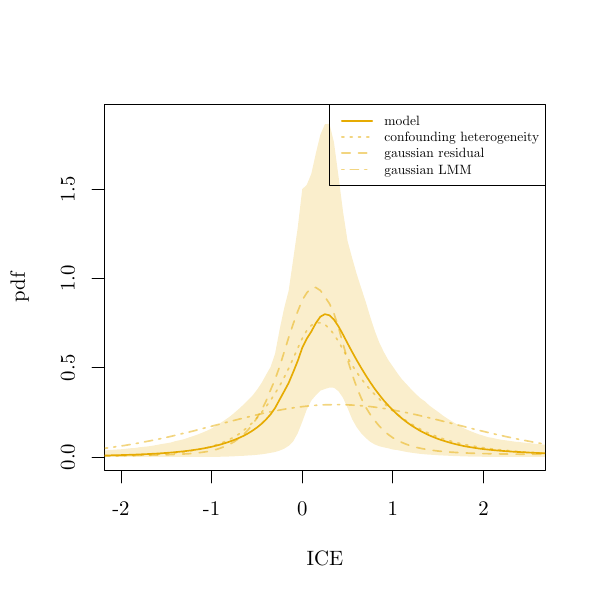} 
	\vspace{-0.5cm}\caption{Posterior distribution of the effect of Hepatic Steatosis on the LVFP using model \eqref{CH3LVmodel3} (solid line) and pointwise $95 \%$ Bayesian credible intervals. Moreover, distributions obtained when either adjusting for confounding-effect heterogeneity, using a Gaussian residual or a Gaussian LMM, respectively, are presented for comparison (see Section \ref{CH3sensa}).}
	\label{CH3fig:icecomparison}
\end{figure}

Thus, model \eqref{CH3LVmodel3} does appropriately describe the observed conditional distributions. The ICE distribution is only identifiable in the absence of other confounders and when the dependence of $Y^{1}-Y^{0}$ and $Y^{0}$ is known. Under assumptions \ref{CH2A2}, \ref{CH2A3}, \ref{CH2A1} and \ref{CH2A4}, the ICE distribution is identified by the distribution of $Z_{1}$ which estimate is presented in Figure \ref{CH3fig:icecomparison} (solid line).  \noindent The ATE has a posterior mean of $0.43$ with a $95 \%$ Bayesian credible interval (BCI) equal to ($0.23, 0.63$). The ATE is thus similar to the effect found by \citet{Chiu2020} and to an ATE estimate of $0.44$, with a $95 \%$ confidence interval of ($0.28,0.60$), obtained using IPTW. The added value of this analysis is inference on other properties of the ICE distribution. For example, $20.6 \%$ (BCI: $8.9 \%, 33.6 \%$) of individuals were estimated to have a harming effect of Hepatic Steatosis on the LVFP that is larger than twice the ATE ($\mathbb{P}(\text{ICE}>2\cdot 0.43)$). The ATE should thus not be used as an individual effect measure.

\subsection{Sensitivity analysis}\label{CH3sensa}

We performed a sensitivity analysis by extending model \eqref{CH3LVmodel3} with confounder-specific residual variances to demonstrate how one could try to adjust for confounding-effect heterogeneity. We have fitted the model \begin{equation}\label{CH3LVmodel2}
\begin{cases}
Y_{i} = \beta_{0} + \boldsymbol{X}_{i}\boldsymbol{\beta}_{X}^T + \left(\boldsymbol{X}_{i}\boldsymbol{\beta_{\text{XA}}}^T+Z_{1i}\right)A_{i}+\delta_{i},\\
\delta_{i}=\varepsilon_{i}+\boldsymbol{\tilde{X}}_{i}\boldsymbol{Z}_{Xi}^T,
\end{cases}
\end{equation} where again $Z_{1i} ~{\sim}~ \text{GM}(\boldsymbol{p}, \boldsymbol{\mu}, \boldsymbol{\tau}^{2})$, with $K{=}5$, $\varepsilon_{i} ~{\sim}~ \text{GM}(\tilde{\boldsymbol{p}}, \boldsymbol{\tilde{\mu}}, \boldsymbol{\tilde{\tau}}^{2})$ with $\tilde{{}K}{=}3$ and $\mathbb{E}[\varepsilon_{i}]=0$,  and now $\boldsymbol{Z}_{Xi}$ are zero-mean Gaussian distributed, while all latent variables are independent. Dichotomized versions of the confounders ($>$ median) are presented by $\boldsymbol{\tilde{X}}_{i}$. This way, one corrects for the shift in mean and the increased variance in the subpopulations with particular values for the confounders. Again, the fit of this distribution to the observed data should be verified. When this model is not appropriate, one could decide to extend the model further by modeling part of the $\boldsymbol{Z}_{X}$ with a mixture of Gaussians instead of the Gaussian distribution. 

The posterior predictive checks in the different exposure and confounder strata can be found in Section \ref{CH3appCH} of the Appendix. The minor differences between the posterior predictive distribution and the observed distribution presented in Section \ref{CH3appNOC} of the Appendix became even smaller. The ATE has a posterior mean of $0.38$ (95$\%$ BCI: $0.19, 0.56$), and the posterior mean of the 
$\mathbb{P}(\text{ICE}>2\cdot 0.43)$ equals $20.4$\% ($8.9$, $34.0$). These estimates were (as expected) very similar to the ones obtained by fitting model \eqref{CH3LVmodel3}, and the estimated distribution is also presented in Figure \ref{CH3fig:icecomparison}. 

We want to emphasize that all latent variables must be modeled with a distribution that is flexible enough to fit the observed outcomes accurately. For comparison, we have also presented the estimated ICE distribution when modeling the residual with a Gaussian distribution (ATE equals $0.42$ ( $0.17$, $0.69$)) and 
$\mathbb{P}(\text{ICE}>2\cdot 0.43)$ equals $12.5$\% ($6.5$, $21.8$). However, in this case, the model for $Y \mid A{=}0$ is misspecified as presented in Figure \ref{CH3fig:fattyliverNOFR} in Appendix \ref{CH3appNOFR}.  When using a Gaussian LMM (ATE equals $0.41$ ($0.17$, $0.65$), and 
$\mathbb{P}(\text{ICE}>2\cdot 0.43)$ equals $36.7$\% ($29.8$, $43.7$)) for the observed outcome in Figure \ref{CH3fig:icecomparison} both without considering confounding effect heterogeneity. In this case, also the model for $Y \mid A{=}1$ seems misspecified as presented in Figure \ref{CH3fig:fattyliverLMM} in Appendix \ref{CH3appLMM}.

\section{Discussion}\label{CH3sec7}
Methods for causal inference have rapidly evolved over the past decades. As a result, making causal claims from observational studies became more common, relying on expert knowledge to back up the untestable identifiability assumption of conditional exchangeability. However, such methods focus on learning (conditional) average treatment effects rather than effect distributions. Therefore, the results are only informative at the level of an individual when (remaining) effect heterogeneity is low. In this work, we have presented an identifiability assumption that suffices to move from (conditional) average treatment effects to quantification of the (conditional) distribution of causal effects. 

In the case of effect heterogeneity, exposure (or absence thereof) increases variability in the outcome as individuals respond differently. When the common causal assumptions \ref{CH2A2}, \ref{CH2A3} and \ref{CH2A1} apply and additionally (conditional) independence of the ICE and the potential outcome under no exposure (Assumption \ref{CH2A4}) can be assumed, inference on the ICE distribution can be drawn from cross-sectional data. The joint distribution of potential outcomes should be linked to the law of observations, which can be learned from the data. Then, deviations from the (C)ATE could be quantified. In case of serious deviations, the shape of the (conditional) ICE distribution can inform about the remaining heterogeneity of the exposure effect. It may illustrate that there is still a severe lack of understanding of the exposure effect as the presence of unmeasured modifiers is considerable. The distribution of the unmeasured modifiers may differ across populations, so estimated ICE distributions could be helpful in the understanding of differences in causal effects \citep{Seamans2021}. 

In contrast to well-established models to estimate CATEs (e.g., ~  \cite{Hahn2020} and \cite{Wager2018}), the focus of the methods presented in this paper is inference on the distribution that quantifies remaining effect heterogeneity. In particular, in settings where limited features are available, the CATEs will not be accurate proxies of the actual ICEs. In the examples presented in this paper, we have used simple linear models  (models \eqref{CH3LVmodel}, \eqref{CH3LVmodel3} and \eqref{CH3LVmodel2}) to estimate the ICE distribution for illustration. However, the causal theory derived in Section \ref{CH3sec2H} applies to more general models of the form presented in \eqref{CH3mixedmodel}. A misspecified mean model will affect the fit of the random exposure effect. The conditional distributions of the data-generating distribution should thus be validated, e.g.~by Bayesian posterior predictive checking as illustrated in the case study. If the posterior distribution for unexposed individuals is off, a more flexible mean model (involving the expected effects of modifiers and confounders) may be necessary. For cases where rich data are available, it will be promising to investigate how flexible machine learning methods can estimate the conditional means in our mixed model \citep{Hajjem2014, Pellagatti2021, Post2024}. For the example and case study, we used a specific Bayesian approach to fit the random-effects model, as it inherently quantifies the uncertainty of the ICE density estimate and simplifies the process by using a one-step approach. However, the reasoning presented in this paper does not rely on the estimation method, and we have listed Bayesian and non-Bayesian alternatives in Section \ref{sec:methods}. What method can be used for a specific application will depend on the available resources, e.g., the Bayesian methods can result in lengthy computation times depending on the complexity of the associational model \eqref{CH3mixedmodel} and the size of the data.

As mentioned before, the fit of the associational model for $F_{Y \mid A=a, \boldsymbol{X}{=}\boldsymbol{x}}$ can be validated when the sample is sufficiently large. On the contrary, the assumptions of conditional exchangeability and conditional independent effect deviation cannot be verified with cross-sectional data. In future work, we will focus on a sensitivity analysis in which models with different dependence structures between $Z_{1}$ and $\varepsilon$ are fitted. As is standard practice for the assumption of conditional exchangeability, assumptions on the (in)dependence of $Y^{1}-Y^{0}$ and $Y^{0}$ should be reviewed by experts since they can seriously affect the estimated ICE distribution. In practice, discussing conditional exchangeability is very challenging as it posits independence between an unobserved quantity (potential outcomes) and an observed one (exposure assignment). Discussing independence between two unobserved quantities ($Y^0$ and the ICE) may be even more challenging and should thus be done with caution. Moreover, assessing the validity of Assumption \ref{CH2A4} is challenging since it is difficult to reason about scale-specific independence. There will be cases where reasoning about the validity of the assumption is impossible, leaving the ICE distribution unidentifiable. Nevertheless, there can be mechanisms for which it is feasible to discuss these assumptions (see, e.g., the clopidogrel example presented by \citet{Post2024}). This work presented the case study to illustrate the methodology only. Therefore, we have not discussed the assumptions \ref{CH2A3} and \ref{CH2A4} with clinical experts. The TBR that is equal to $\mathbb{P}(Y^{1}-Y^{0}<0)$, was under these assumptions estimated to equal $0.26$ (95 $\%$ BCI: $0.08, 0.40$), while it seems biologically impossible that there exist individuals in which LVFP decreases due to Hepatic Steatosis. The non-zero probability may result from variability in repeated measurements of an individual's LVFP. The potential outcome of individual $i$ for exposure level $a$ at time $1$, $Y^{a}_{i1}$, could differ from that at time $2$, $Y^{a}_{i2}$. Then, $N_{Y}$ represents both between individual and within (over multiple repeats) variability of $Y^{0}$, so that $\mathbb{P}(Y^{0}_{i1}-Y^{0}_{i2}<0)$ as well as $\mathbb{P}(Y^{1}_{i1} - Y^{0}_{i2}<0)$ can indeed be positive. Finally, the heterogeneity in effect might be overestimated when $Y^1-Y^0$ and $Y^0$ are in reality positively related, e.g., when $U_{1} = \theta_{U}N_{Y} + U_{2}$ for $\theta_{U}>0$ and $U_{2} \independent N_{Y} \mid \boldsymbol{X}$. In that case, the actual ICE distribution will be less variable than the distribution of $Z_{1}$, and the estimated $\mathbb{P}(Y^1-Y^{0}<0)$ will decrease. These types of discussions should take place with experts in the field. We, therefore, advocate that assumptions on the (conditional) joint distribution of potential outcomes should become part of the discussions with experts to boost research on the heterogeneity of treatment (or exposure) effects.


\section*{Acknowledgments}
We sincerely thank Michelle Long and Alison Pedley for sharing the \texttt{SAS} script to reproduce the sub-sample of the Framingham Heart Study as used in Section \ref{CH3sec6}. The authors thank the associate editor, and three anonymous reviewers for their valuable comments. 

\section*{Author contributions}
RAJP conceptualized the study, developed the methodology, implemented the simulation code, conducted the case study, and drafted the original manuscript. ERvdH supervised the project and contributed to the review and editing of the manuscript. Both authors read and approved the final manuscript.

\section*{Funding information}
Authors state no funding involved.

\section*{Conflict of interest}
Authors state no conflict of interest.

\section*{Data availability statement}
The \texttt{SAS} and \texttt{R} codes used for simulation and analysis of the example presented in section \ref{CH3sec:bayes} and \ref{CH3confhet} and for the analysis in Section \ref{CH3sec6} can be found at \url{https://github.com/RAJP93/ICE-distribution}.


\begin{thebibliography}{54}
\providecommand{\natexlab}[1]{#1}
\providecommand{\url}[1]{\texttt{#1}}
\expandafter\ifx\csname urlstyle\endcsname\relax
  \providecommand{\doi}[1]{doi: #1}\else
  \providecommand{\doi}{doi: \begingroup \urlstyle{rm}\Url}\fi
\bibitem[Hand(1992)]{Hand1992}
David~J. Hand.
\newblock {On comparing two treatments}.
\newblock \emph{The American Statistician}, 46\penalty0 (3):\penalty0 190--192, 1992.
\newblock \doi{10.1080/00031305.1992.10475881}.

\bibitem[Greenland et~al.(2019)Greenland, Fay, Brittain, Shih, Follmann, Gabriel, and Robins]{Greenland2019}
Sander Greenland, Michael~P. Fay, Erica~H. Brittain, Joanna~H. Shih, Dean~A. Follmann, Erin~E. Gabriel, and James~M. Robins.
\newblock {On Causal Inferences for Personalized Medicine: How Hidden Causal Assumptions Led to Erroneous Causal Claims About the D-Value}.
\newblock \emph{The American Statistician}, 74\penalty0 (3):\penalty0 243--248, 2019.
\newblock \doi{10.1080/00031305.2019.1575771}.

\bibitem[Holland(1986)]{Holland1986}
Paul~W. Holland.
\newblock {Statistics and causal inference}.
\newblock \emph{Journal of the American Statistical Association}, 81\penalty0 (396):\penalty0 945--960, 1986.
\newblock \doi{10.1080/01621459.1986.10478354}.

\bibitem[Kennedy et~al.(2023)Kennedy, Balakrishnan, and Wasserman]{Kennedy2023}
E~H Kennedy, S~Balakrishnan, and L~A Wasserman.
\newblock {Semiparametric counterfactual density estimation}.
\newblock \emph{Biometrika}, 110\penalty0 (4):\penalty0 875--896, 03 2023.
\newblock \doi{10.1093/biomet/asad017}.

\bibitem[Robins et~al.(2000)Robins, Hern{\'{a}}n, and Brumback]{Robins2000}
James~M. Robins, Miguel~A Hern{\'{a}}n, and Babette Brumback.
\newblock {Marginal structural models and causal inference in epidemiology}.
\newblock \emph{Epidemiology}, 11\penalty0 (5):\penalty0 550--560, 2000.
\newblock \doi{10.1097/00001648-200009000-00011}.

\bibitem[Caron et~al.(2022)Caron, Baio, and Manolopoulou]{Caron2022}
Alberto Caron, Gianluca Baio, and Ioanna Manolopoulou.
\newblock Estimating individual treatment effects using non-parametric regression models: A review.
\newblock \emph{Journal of the Royal Statistical Society: Series A (Statistics in Society)}, 185\penalty0 (3):\penalty0 1115--1149, 2022.
\newblock \doi{10.1111/rssa.12824}.

\bibitem[Lu et~al.(2018)Lu, Sadiq, Feaster, and Ishwaran]{Lu2018}
Min Lu, Saad Sadiq, Daniel~J. Feaster, and Hemant Ishwaran.
\newblock {Estimating Individual Treatment Effect in Observational Data Using Random Forest Methods}.
\newblock \emph{Journal of Computational and Graphical Statistics}, 27\penalty0 (1):\penalty0 209--219, 2018.
\newblock \doi{10.1080/10618600.2017.1356325}.


\bibitem[Post et~al.(2024)Post, Petkovic, {van den Heuvel}, and {van den Heuvel}]{Post2024}
Richard A~J Post, Marko Petkovic, Isabel~L {van den Heuvel}, and Edwin~R {van den Heuvel}.
\newblock {Flexible Machine Learning Estimation of Conditional Average Treatment Effects: A Blessing and a Curse}.
\newblock \emph{Epidemiology}, 35\penalty0 (1), 2024.
\newblock \doi{10.1097/EDE.0000000000001684}

\bibitem[Pearl(1999)]{Pearl1999}
Judea Pearl.
\newblock {Probabilities Of Causation: Three Counterfactual Interpretations And Their Identification}.
\newblock \emph{Synthese}, 121\penalty0 (1):\penalty0 93--149, 1999.
\newblock \doi{10.1023/A:1005233831499}.

\bibitem[Tian and Pearl(2000)]{Tian2000}
Jin Tian and Judea Pearl.
\newblock {Probabilities of causation: Bounds and identification}.
\newblock \emph{Annals of Mathematics and Artificial Intelligence}, 28\penalty0 (1):\penalty0 287--313, 2000.
\newblock \doi{10.1023/A:1018912507879}.

\bibitem[Mueller and Pearl(2023)]{Mueller2023}
Scott Mueller and Judea Pearl.
\newblock Personalized decision making – a conceptual introduction.
\newblock \emph{Journal of Causal Inference}, 11\penalty0 (1):\penalty0 20220050, 2023.
\newblock \doi{10.1515/jci-2022-0050}.

\bibitem[Li and Pearl(2022)]{Li2022}
Ang Li and Judea Pearl.
\newblock Probabilities of causation with nonbinary treatment and effect.
\newblock arxiv, 2022.
\newblock \doi{10.48550/arXiv.2208.09568}

\bibitem[Hansen(2008)]{Hansen2008}
Ben~B. Hansen.
\newblock {The prognostic analogue of the propensity score}.
\newblock \emph{Biometrika}, 95\penalty0 (2):\penalty0 481--488, 2008.
\newblock \doi{10.1093/biomet/asn004}.

\bibitem[Hahn et~al.(2020)Hahn, Jared, and Carvalho]{Hahn2020}
Richard~P. Hahn, Jared~S. Murray, and Carlos~M. Carvalho.
\newblock {Bayesian regression tree models for causal inference: Regularization, confounding, and heterogeneous effects (with discussion)}
\newblock \emph{Bayesian Analysis}, 15\penalty0 (3):\penalty0 965--1056, 2020.
\newblock \doi{10.1214/19-BA1195}.

\bibitem[Huang et~al.(2016)Huang, Fang, Hanley, and Rosenblum]{Huang2017}
Emily~J. Huang, Ethan~X. Fang, Daniel~F. Hanley, and Michael Rosenblum.
\newblock {Inequality in treatment benefits: Can we determine if a new treatment benefits the many or the few?}
\newblock \emph{Biostatistics}, 18\penalty0 (2):\penalty0 308--324, 11 2016.
\newblock \doi{10.1093/biostatistics/kxw049}.

\bibitem[Huang et~al.(2019)Huang, Fang, Hanley, and Rosenblum]{Huang2019}
Emily~J. Huang, Ethan~X. Fang, Daniel~F. Hanley, and Michael Rosenblum.
\newblock Constructing a confidence interval for the fraction who benefit from treatment, using randomized trial data.
\newblock \emph{Biometrics}, 75\penalty0 (4):\penalty0 1228--1239, 2019.
\newblock \doi{10.1111/biom.13101}.

\bibitem[Su and Li(2023)]{Su2023}
Yongchang Su and Xinran Li.
\newblock {Treatment effect quantiles in stratified randomized experiments and matched observational studies}.
\newblock \emph{Biometrika}, page asad030, 05 2023.
\newblock \doi{10.1093/biomet/asad030}.

\bibitem[Lei and Cand\`{e}s(2021)]{Lei2020}
Lihua Lei and Emmanuel~J. Cand\`{e}s.
\newblock Conformal inference of counterfactuals and individual treatment effects.
\newblock \emph{Journal of the Royal Statistical Society: Series B (Statistical Methodology)}, 83\penalty0 (5):\penalty0 911--938, 2021.
\newblock \doi{10.1111/rssb.12445}.

\bibitem[Jin et~al.(2023)Jin, Ren, and Candès]{Jin2022}
Ying Jin, Zhimei Ren, and Emmanuel~J. Candès.
\newblock Sensitivity analysis of individual treatment effects: A robust conformal inference approach.
\newblock \emph{Proceedings of the National Academy of Sciences}, 120\penalty0 (6):\penalty0 e2214889120, 2023.
\newblock \doi{10.1073/pnas.2214889120}.

\bibitem[Mingzhang~Yin and Blei(2022)]{Yin2022}
Yixin~Wang Mingzhang~Yin, Claudia~Shi and David~M. Blei.
\newblock Conformal sensitivity analysis for individual treatment effects.
\newblock \emph{Journal of the American Statistical Association}, 0\penalty0 (0):\penalty0 1--14, 2022.
\newblock \doi{10.1080/01621459.2022.2102503}.

\bibitem[Chernozhukov et~al.(2023)Chernozhukov, Wüthrich, and Zhu]{Chernozhukov2023}
Victor Chernozhukov, Kaspar Wüthrich, and Yinchu Zhu.
\newblock Toward personalized inference on individual treatment effects.
\newblock \emph{Proceedings of the National Academy of Sciences}, 120\penalty0 (7):\penalty0 e2300458120, 2023.
\newblock \doi{10.1073/pnas.2300458120}.

\bibitem[Chernozhukov et~al.(2021)Chernozhukov, Wüthrich, and Zhu]{Chernozhukov2021}
Victor Chernozhukov, Kaspar Wüthrich, and Yinchu Zhu.
\newblock Distributional conformal prediction.
\newblock \emph{Proceedings of the National Academy of Sciences}, 118\penalty0 (48):\penalty0 e2107794118, 2021.
\newblock \doi{10.1073/pnas.2107794118}.Zhang2013

\bibitem[Zhang et~al.(2013)Zhang, Wang, Nie, and Soon]{Zhang2013}
Zhiwei Zhang, Chenguang Wang, Lei Nie, and Guoxing Soon.
\newblock {Assessing the heterogeneity of treatment effects via potential outcomes of individual patients}.
\newblock \emph{Journal of the Royal Statistical Society: Series C (Applied Statistics)}, 62\penalty0 (5):\penalty0 687--704, 2013.
\newblock \doi{10.1111/rssc.12012}.

\bibitem[Yin et~al.(2018)Yin, Liu, and Geng]{Yin2018}
Yunjian Yin, Lan Liu, and Zhi Geng.
\newblock {Assessing the treatment effect heterogeneity with a latent variable}.
\newblock \emph{Statistica Sinica}, 28\penalty0 (1):\penalty0 115--135, 2018.

\bibitem[Laubender et~al.(2020)Laubender, Mansmann, and Lauseker]{Laubender2020}
Ruediger~P. Laubender, Ulrich Mansmann, and Michael Lauseker.
\newblock Estimating the distribution of heterogeneous treatment effects from treatment responses and from a predictive biomarker in a parallel-group rct: A structural model approach.
\newblock \emph{Biometrical Journal}, 62\penalty0 (3):\penalty0 697--711, 2020.
\newblock \doi{10.1002/bimj.201800370}.

\bibitem[Shahn and Madigan(2017)]{Shahn2017}
Zach Shahn and David Madigan.
\newblock {Latent class mixture models of treatment effect heterogeneity}.
\newblock \emph{Bayesian Analysis}, 12\penalty0 (3):\penalty0 831--854, 2017.
\newblock \doi{10.1214/16-BA1022}.

\bibitem[Neyman(1923)]{Neyman1990}
Jerzy Neyman.
\newblock {On the Application of Probability Theory to Agricultural Experiments. Essay on Principles}.
\newblock \emph{Statistical Science}, 5\penalty0 (4):\penalty0 465--472, 1923.
\newblock \doi{10.1214/ss/1177012031}.

\bibitem[Rubin(1974)]{Rubin1974}
Donald~B. Rubin.
\newblock {Estimating causal effects of treatments in randomized and nonrandomized studies}.
\newblock \emph{Journal of Educational Psychology}, 66\penalty0 (5):\penalty0 688--701, 1974.
\newblock \doi{10.1037/h0037350}.

\bibitem[Robins and Greenland(1989)]{Robins1989}
James Robins and Sander Greenland.
\newblock The probability of causation under a stochastic model for individual risk.
\newblock \emph{Biometrics}, 45\penalty0 (4):\penalty0 1125--1138, 1989.
\newblock \doi{10.2307/2531765}.

\bibitem[VanderWeele and Robins(2012)]{VanderWeele2012}
Tyler~J. VanderWeele and James~M. Robins.
\newblock Stochastic counterfactuals and stochastic sufficient causes.
\newblock \emph{Statistica Sinica}, 22\penalty0 (1):\penalty0 379--392, 2012.
\newblock \doi{10.5705/ss.2008.186}

\bibitem[Wager and Athey(2018)]{Wager2018}
Stefan Wager and Susan Athey.
\newblock {Estimation and Inference of Heterogeneous Treatment Effects using Random Forests}.
\newblock \emph{Journal of the American Statistical Association}, 113\penalty0 (523):\penalty0 1228--1242, 2018.
\newblock \doi{10.1080/01621459.2017.1319839}

\bibitem[Hern{\'{a}}n and Robins(2020)]{Hernan2019}
Miguel~A Hern{\'{a}}n and James~M. Robins.
\newblock \emph{Causal Inference: What If.}
\newblock Boca Raton: Chapman {\&} Hall/CRC, Boca Raton, Florida, 1st edition, 2020.
\newblock URL \url{https://www.hsph.harvard.edu/miguel-hernan/causal-inference-book/}.

\bibitem[Cole and Frangakis(2009)]{Cole2009}
Stephen~R. Cole and Constantine~E. Frangakis.
\newblock {The Consistency Statement in Causal Inference: A Definition or an Assumption?}
\newblock \emph{Epidemiology}, 20\penalty0 (1), 2009.
\newblock \doi{10.1097/EDE.0b013e31818ef366}.

\bibitem[Pearl(2009)]{Pearl2009book}
Judea Pearl.
\newblock \emph{{Causality: Models, reasoning, and inference}}.
\newblock Cambridge University Press, 2nd edition, 2009.
\newblock ISBN 9780511803161.
\newblock \doi{10.1017/CBO9780511803161}.

\bibitem[Imbens and Rubin (2015)]{Imbens2015}
Guido~W. Imbens and Donald B. Rubin.
\newblock \emph{{Causal Inference for Statistics, Social, and Biomedical Sciences: An Introduction}}.
\newblock Cambridge University Press, 1st edition, 2015.
\newblock ISBN 9781139025751.
\newblock \doi{10.1017/CBO9781139025751}.


\bibitem[Peters et~al.(2018)Peters, Janzing, and Sch{\"o}lkopf]{Peters2017}
Jonas Peters, Dominik Janzing, and Bernhard Sch{\"o}lkopf.
\newblock \emph{{Elements of causal inference: foundations and learning algorithms}}.
\newblock The MIT Press, Cambridge, 1st edition, 2018.
\newblock ISBN 9780262037310.
\newblock \doi{10.1080/00949655.2018.1505197}.

\bibitem[Munafò et~al.(2017)Munafò, Tilling, Taylor, Evans, and Davey~Smith]{Munafo2017}
Marcus~R Munafò, Kate Tilling, Amy~E Taylor, David~M Evans, and George Davey~Smith.
\newblock {Collider scope: when selection bias can substantially influence observed associations}.
\newblock \emph{International Journal of Epidemiology}, 47\penalty0 (1):\penalty0 226--235, 09 2017.
\newblock \doi{10.1093/ije/dyx206}.

\bibitem[Meister(2009)]{Meister2009}
Alexander Meister.
\newblock \emph{Density Deconvolution}, pages 5--105.
\newblock Springer Berlin Heidelberg, Berlin, Heidelberg, 2009.
\newblock ISBN 978-3-540-87557-4.
\newblock \doi{10.1007/978-3-540-87557-4_2}.

\bibitem[Bonvini and Kennedy(2021)]{Bonvini2020}
Matteo Bonvini and Edward~H. Kennedy.
\newblock Sensitivity analysis via the proportion of unmeasured confounding.
\newblock \emph{Journal of the American Statistical Association}, 117\penalty0 (539):\penalty0 1--11, 2021.
\newblock \doi{10.1080/01621459.2020.1864382}.

\bibitem[Browne and Draper(2006)]{Browne2006}
William~J. Browne and David Draper.
\newblock {A comparison of Bayesian and likelihood-based methods for fitting multilevel models}.
\newblock \emph{Bayesian Analysis}, 1\penalty0 (3):\penalty0 473--514, 2006.
\newblock \doi{10.1214/06-BA117}.

\bibitem[McCulloch and Neuhaus(2011)]{McCulloch2011}
Charles~E. McCulloch and John~M. Neuhaus.
\newblock {Misspecifying the Shape of a Random Effects Distribution: Why Getting It Wrong May Not Matter}.
\newblock \emph{Statistical Science}, 26\penalty0 (3):\penalty0 388 -- 402, 2011.
\newblock \doi{10.1214/11-STS361}.

\bibitem[Verbeke and Lesaffre(1996)]{Verbeke1996}
Geert Verbeke and Emmanuel Lesaffre.
\newblock {A Linear Mixed-Effects Model With Heterogeneity in the Random-Effects Population}.
\newblock \emph{Journal of the American Statistical Association}, 91\penalty0 (433):\penalty0 217--221, 1996.
\newblock \doi{10.2307/2291398}.

\bibitem[Proust and Jacqmin-Gadda(2005)]{Proust2005}
C{\'{e}}cile Proust and H{\'{e}}l{\`{e}}ne Jacqmin-Gadda.
\newblock {Estimation of linear mixed models with a mixture of distribution for the random effects}.
\newblock \emph{Comput. Methods Programs Biomed.}, 78\penalty0 (2):\penalty0 165--173, 2005.
\newblock \doi{10.1016/j.cmpb.2004.12.004}.

\bibitem[Zhang and Davidian(2001)]{Zhang2001}
Daowen Zhang and Marie Davidian.
\newblock {Linear Mixed Models with Flexible Distributions of Random Effects for Longitudinal Data}.
\newblock \emph{Biometrics}, 57\penalty0 (3):\penalty0 795--802, 2001.
\newblock \doi{10.1111/j.0006-341X.2001.00795.x}.

\bibitem[Kleinman and Ibrahim(1998)]{Kleinman1998}
Ken~P. Kleinman and Joseph~G. Ibrahim.
\newblock {A Semiparametric Bayesian Approach to the Random Effects Model}.
\newblock \emph{Biometrics}, 54\penalty0 (3):\penalty0 921, 1998.
\newblock \doi{10.2307/2533846}.

\bibitem[Gelman et~al.(2021)Gelman, Carlin, Stern, Dunson, Vehtari, and Rubin]{Gelman2021}
Andrew Gelman, John~B. Carlin, Hal~S. Stern, David~B. Dunson, Aki Vehtari, and Donald~B. Rubin.
\newblock \emph{{Bayesian data analysis}}.
\newblock CRC Press, third edition, 2021.
\newblock ISBN 9781439898208.

\bibitem[Ho and Hu(2008)]{Ho2008}
Remus~K.W. Ho and Inchi Hu.
\newblock {Flexible modelling of random effects in linear mixed models - A Bayesian approach}.
\newblock \emph{Computational Statistics \& Data Analysis}, 52\penalty0 (3):\penalty0 1347--1361, 2008.
\newblock \doi{10.1016/j.csda.2007.09.005}.

\bibitem[Rousseau and Mengersen(2011)]{Rousseau2011}
Judith Rousseau and Kerrie Mengersen.
\newblock {Asymptotic behaviour of the posterior distribution in overfitted mixture models}.
\newblock \emph{Journal of the Royal Statistical Society. Series B (Methodological)}, 73\penalty0 (5):\penalty0 689--710, 2011.
\newblock \doi{10.1111/j.1467-9868.2011.00781.x}.

\bibitem[Dunson(2009)]{Dunson2009}
David~B. Dunson.
\newblock Bayesian nonparametric hierarchical modeling.
\newblock \emph{Biometrical Journal}, 51\penalty0 (2):\penalty0 273--284, 2009.
\newblock \doi{10.1002/bimj.200800183}.

\bibitem[Ohlssen et~al.(2007)Ohlssen, Sharples, and Spiegelhalter]{Ohlssen2007}
D.~I. Ohlssen, L.~D. Sharples, and D.~J. Spiegelhalter.
\newblock Flexible random-effects models using bayesian semi-parametric models: applications to institutional comparisons.
\newblock \emph{Statistics in Medicine}, 26\penalty0 (9):\penalty0 2088--2112, 2007.
\newblock \doi{10.1002/sim.2666}.

\bibitem[Gelman(2006)]{Gelman2006}
Andrew Gelman.
\newblock {Prior distributions for variance parameters in hierarchical models (Comment on Article by Browne and Draper)}.
\newblock \emph{Bayesian Analysis}, 1\penalty0 (3):\penalty0 515--534, 2006.
\newblock \doi{10.1214/06-BA117A}.

\bibitem[Chiu et~al.(2020)Chiu, Pedley, Massaro, Benjamin, Mitchell, McManus, Aragam, Vasan, Cheng, and Long]{Chiu2020}
Laura~S. Chiu, Alison Pedley, Joseph~M. Massaro, Emelia~J. Benjamin, Gary~F. Mitchell, David~D. McManus, Jayashri Aragam, Ramachandran~S. Vasan, Susan Cheng, and Michelle~T. Long.
\newblock The association of non-alcoholic fatty liver disease and cardiac structure and function—framingham heart study.
\newblock \emph{Liver International}, 40\penalty0 (10):\penalty0 2445--2454, 2020.
\newblock \doi{10.1111/liv.14600}.

\bibitem[Ghidey et~al.(2010)Ghidey, Lesaffre, and Verbeke]{Ghidey2010}
Wendimagegn Ghidey, Emmanuel Lesaffre, and Geert Verbeke.
\newblock {A comparison of methods for estimating the random effects distribution of a linear mixed model}.
\newblock \emph{Statistical Methods in Medical Research}, 19\penalty0 (6):\penalty0 575--600, 2010.
\newblock \doi{10.1177/0962280208091686}.

\bibitem[Rubio and Steel(2018)]{Rubio2018}
F.~J. Rubio and M.~F.J. Steel.
\newblock {Flexible linear mixed models with improper priors for longitudinal and survival data}.
\newblock \emph{Electronic Journal of Statistics}, 12\penalty0 (1):\penalty0 572--598, 2018.
\newblock \doi{10.1214/18-EJS1401}.

\bibitem[Vehtari et~al.(2021)Vehtari, Gelman, Simpson, Carpenter, and B\"{u}rkner]{Vehtari2021}
Aki Vehtari, Andrew Gelman, Daniel Simpson, Bob Carpenter, and Paul-Christian B\"{u}rkner.
\newblock {Rank-Normalization, Folding, and Localization: An Improved $\widehat{R}$ for Assessing Convergence of MCMC (with Discussion)}.
\newblock \emph{Bayesian Analysis}, 16\penalty0 (2):\penalty0 667 -- 718, 2021.
\newblock \doi{10.1214/20-BA1221}.

\bibitem[Seamans et~al.(2021)Seamans, Hong, Ackerman, Schmid, and Stuart]{Seamans2021}
Marissa~J. Seamans, Hwanhee Hong, Benjamin Ackerman, Ian Schmid, and Elizabeth~A. Stuart.
\newblock {Generalizability of subgroup effects}.
\newblock \emph{Epidemiology}, 32\penalty0 (3):\penalty0 389--392, 2021.
\newblock \doi{10.1097/EDE.0000000000001329}.

\bibitem[Hajjem et~al.(2014)Hajjem, Bellavance, and Larocque]{Hajjem2014}
Ahlem Hajjem, François Bellavance, and Denis Larocque.
\newblock Mixed-effects random forest for clustered data.
\newblock \emph{Journal of Statistical Computation and Simulation}, 84\penalty0 (6):\penalty0 1313--1328, 2014.
\newblock \doi{10.1080/00949655.2012.741599}.

\bibitem[Pellagatti et~al.(2021)Pellagatti, Masci, Ieva, and Paganoni]{Pellagatti2021}
Massimo Pellagatti, Chiara Masci, Francesca Ieva, and Anna~M. Paganoni.
\newblock Generalized mixed-effects random forest: A flexible approach to predict university student dropout.
\newblock \emph{Statistical Analysis and Data Mining: The ASA Data Science Journal}, 14\penalty0 (3):\penalty0 241--257, 2021.
\newblock \doi{10.1002/sam.11505}.

\bibitem[Zhang et~al.(2024)Zhang, Geng, Li, and Ding]{Zhang2024}
Chao Zhang, Zhi Geng, Wei Li, and Peng Ding.
\newblock Identifying and bounding the probability of necessity for causes of effects with ordinal outcomes.
\newblock \emph{arXiv preprint}, 2024.
\newblock \url{https://arxiv.org/abs/2411.01234}.

\bibitem[Gadbury et~al.(2004)Gadbury, Iyer, and Albert]{Gadbury2004}
Gary L. Gadbury, Hari K. Iyer, and Jeffrey M. Albert.
\newblock Individual treatment effects in randomized trials with binary outcomes.
\newblock \emph{Journal of Statistical Planning and Inference}, 121\penalty0 (2):\penalty0 163--174, 2004.
\newblock \doi{10.1016/S0378-3758(03)00115-0}.

\bibitem[Wu et~al.(2024)Wu, Ding, Geng, and Liu]{Wu2024}
Peng Wu, Peng Ding, Zhi Geng, and Yue Liu.
\newblock Quantifying individual risk for binary outcome.
\newblock \emph{arXiv preprint}, 2024.
\newblock \url{https://arxiv.org/abs/2402.10537}.

\bibitem[Kallus(2022)]{Kallus2022}
Nathan Kallus.
\newblock What's the harm? Sharp bounds on the fraction negatively affected by treatment.
\newblock \emph{arXiv preprint}, 2022.
\newblock \url{https://arxiv.org/abs/2205.10327}.

\bibitem[Ben-Michael et~al.(2024)Ben-Michael, Imai, and Jiang]{Ben-Michael2024}
Eli Ben-Michael, Kosuke Imai, and Zhichao Jiang.
\newblock Policy learning with asymmetric counterfactual utilities.
\newblock \emph{Journal of the American Statistical Association}, 119\penalty0 (548):\penalty0 3045--3058, 2024.
\newblock \doi{10.1080/01621459.2023.2300507}.

\bibitem[Sarvet and Stensrud(2023)]{Sarvet2023}
Aaron L. Sarvet and Mats J. Stensrud.
\newblock Perspective on ‘Harm’ in personalized medicine.
\newblock \emph{American Journal of Epidemiology}, 2023.
\newblock \doi{10.1093/aje/kwad162}.

\bibitem[Lu et~al.(2022)Lu, Geng, Li, Zhu, and Jia]{Lu2022}
Zitong Lu, Zhi Geng, Wei Li, Shengyu Zhu, and Jinzhu Jia.
\newblock Evaluating causes of effects by posterior effects of causes.
\newblock \emph{Biometrika}, 110\penalty0 (2):\penalty0 449--465, 2022.
\newblock \doi{10.1093/biomet/asac038}.

\bibitem[Li et~al.(2023)Li, Zheng, Cao, Geng, Liu, and Wu]{Li2023}
Haoxuan Li, Chunyuan Zheng, Yixiao Cao, Zhi Geng, Yue Liu, and Peng Wu.
\newblock Trustworthy policy learning under the counterfactual no-harm criterion.
\newblock In \emph{Proceedings of the 40th International Conference on Machine Learning}, volume 202 of \emph{Proceedings of Machine Learning Research}, pages 20575--20598. PMLR, July 2023.
\newblock \url{https://proceedings.mlr.press/v202/li23ay.html}.


\end{thebibliography}

\newpage
\section*{Appendix}
\appendix






\section{Proof of Proposition \ref{prop1}}
If $Y^{1}-Y^{0} \mid \boldsymbol{X}=\boldsymbol{x}$ is degenerate then $Y^{1}_{i} = Y^{0}_i + f_{Y1}(\boldsymbol{X}_{i})$ for some functional $f_{Y1}$. Therefore,
\begin{equation}
    \text{var}(Y^{1} \mid \boldsymbol{X} = \boldsymbol{x}) = \text{var}(Y^{0} +f_{Y1}(\boldsymbol{X}) \mid \boldsymbol{X} = \boldsymbol{x}) = \text{var}(Y^{0}\mid \boldsymbol{X} = \boldsymbol{x}).
\end{equation} By assumption \ref{CH2A3}, $\text{var}(Y^{a} \mid \boldsymbol{X}=\boldsymbol{x}) = \text{var}(Y^{a} \mid \boldsymbol{X}=\boldsymbol{x}, A=a)$, which is equal to $\text{var}(Y \mid \boldsymbol{X}=\boldsymbol{x}, A=a)$ by Assumption \ref{CH2A2} and is well defined by Assumption \ref{CH2A1}. As a result, 
\begin{equation}
    \text{var}(Y \mid \boldsymbol{X} = \boldsymbol{x}, A=1) = \text{var}(Y\mid \boldsymbol{X} = \boldsymbol{x}, A=0).
\end{equation}\qed

\section{Proof of Proposition \ref{prop2}}
By assumption \ref{CH2A2}, \ref{CH2A1} and \ref{CH2A3}, $\text{var}(Y^{a} \mid \boldsymbol{X}=\boldsymbol{x}) = \text{var}(Y \mid \boldsymbol{X}=\boldsymbol{x}, A=a), $ so that 
\begin{equation}
    \text{var}(Y^{1} \mid \boldsymbol{X} = \boldsymbol{x}, A=1) > \text{var}(Y^{0}\mid \boldsymbol{X} = \boldsymbol{x}, A=0).
\end{equation} If $Y^1-Y^0 \mid \boldsymbol{X}=\boldsymbol{x}$ would be non-degenerate, as shown in the proof of Proposition \ref{prop1},  $\text{var}(Y^{1} \mid \boldsymbol{X} = \boldsymbol{x}, A=1) = \text{var}(Y^{0}\mid \boldsymbol{X} = \boldsymbol{x}, A=0)$, contradicting the variance inequality, thus completing the proof by contradiction. \qed

\section{Proof of Lemma \ref{CH5th2}}
The conditional ICE distribution, $\mathbb{P}(Y^{1}-Y^{0} \leq y ~{\mid}~ \boldsymbol{X}{=}\boldsymbol{x})$, equals 
\begin{equation}
\mathbb{P}(Y^{1}-Y^{0}\leq y \mid \boldsymbol{X}{=}\boldsymbol{x})  =\int_{y_{1}=-\infty}^{\infty} \int_{y_{2}=y_{1}-y}^{\infty}  1 dF_{Y^{1},Y^{0}\mid \boldsymbol{X}{=}\boldsymbol{x}}(y_{1},y_{2}),  
\end{equation}  where $\int g(x) dF_{X}(x)$ is the Lebesque-Stieltjes integral of $g(X)$ with respect to probability law $F_{X}$. For the parameterization in SCM \eqref{CH3SCMsurv2},  by the law of total probability, this can be written as
\begin{equation}
 \int_{n_{Y}=-\infty}^{\infty} \int_{y_{1}=-\infty}^{\infty} \int_{y_{2}=y_{1}-y}^{\infty}   1 dF_{Y^{1},Y^{0} \mid N_{Y}{=}n_{Y}, \boldsymbol{X}{=}\boldsymbol{x}}(y_{1},y_{2}) dF_{N_{Y}\mid \boldsymbol{X}{=}\boldsymbol{x}}(n_{Y}).
\end{equation} As $Y^{0}\mid  N_{Y}{=}n_{Y}, \boldsymbol{X}{=}\boldsymbol{x}$ is a degenerate random variable $Y^{0}\mid  N_{Y}{=}n_{Y}, \boldsymbol{X}{=}\boldsymbol{x}$ is as well, so that
$Y^{1} \independent Y^{0} ~{\mid}~ \boldsymbol{X}{=}\boldsymbol{x}, N_{Y}{=}n_{Y}$, and thus $\mathbb{P}(Y^{1}-Y^{0}\leq y ~{\mid}~ \boldsymbol{X}{=}\boldsymbol{x})$ equals
\begin{equation}
	 \int_{n_{Y}=-\infty}^{\infty} \int_{y_{1}=-\infty}^{\infty} \int_{y_{2}=y_{1}-y}^{\infty}  1 dF_{Y^{1}\mid  N_{Y}{=}n_{Y}, \boldsymbol{X}{=}\boldsymbol{x}}(y_{1}) dF_{Y^{0}\mid N_{Y}{=}n_{Y}, \boldsymbol{X}{=}\boldsymbol{x}}(y_{2}) dF_{N_{Y} \mid \boldsymbol{X}{=}\boldsymbol{x}}(n_{Y})
\end{equation} Since $Y^{1}, Y^{0} \independent A ~{\mid}~ \boldsymbol{X}$ (and $Y^{1}, Y^{0} \independent A ~{\mid}~ \boldsymbol{X}$), $\mathbb{P}(Y^{1}-Y^{0}\leq y ~{\mid}~ \boldsymbol{X})$ equals
\begin{equation}
	\int_{n_{Y}=-\infty}^{\infty}\int_{y_{1}=-\infty}^{\infty} \int_{y_{2}=y_{1}-y}^{\infty}  1 dF_{Y^{1}\mid  A{=}1,N_{Y}{=}n_{Y}, \boldsymbol{X}{=}\boldsymbol{x}}(y_{1}) dF_{Y^{0}\mid A{=}0,N_{Y}{=}n_{Y}, \boldsymbol{X}{=}\boldsymbol{x}}(y_{2})dF_{N_{Y} \mid \boldsymbol{X}{=}\boldsymbol{x}}(n_{Y}).
\end{equation}
Finally, by causal consistency and positivity, $\mathbb{P}(Y^{1}-Y^{0}\leq y ~{\mid}~ \boldsymbol{X})$ equals
\begin{equation}
\int_{n_{Y}=-\infty}^{\infty}\int_{y_{1}=-\infty}^{\infty} \int_{y_{2}=y_{1}-y}^{\infty}  1 dF_{Y\mid  A{=}1, N_{Y}{=}n_{Y}, \boldsymbol{X}{=}\boldsymbol{x}}(y_{1}) dF_{Y\mid A{=}0,  N_{Y}{=}n_{Y}, \boldsymbol{X}{=}\boldsymbol{x}}(y_{2}) dF_{N_{Y} \mid \boldsymbol{X}{=}\boldsymbol{x}}(n_{Y}).
\end{equation}\qed

\section{Proof of Proposition \ref{Cor1}}
By definition of SCM \eqref{CH3SCMsurv2}, $$\mathbb{P}(Y^{1}-Y^{0} \leq y ~{\mid}~ \boldsymbol{X}{=}\boldsymbol{x}) = \mathbb{P}(G_{1} \leq y),$$ where $G_{1}$ is equal in distribution to $(f_{Y1}(\boldsymbol{X}) + U_{1}) \mid \boldsymbol{X}{=}\boldsymbol{x}$. By conditional exchangeability, $\boldsymbol{X}$ is contained in $\boldsymbol{X}$ so that $f_{Y1}(\boldsymbol{X})\mid \boldsymbol{X}{=}\boldsymbol{x}$ is degenerate. Moreover, under Assumption \ref{CH2A4}, $U_{1} \independent Y^{0} \mid \boldsymbol{X}{=}\boldsymbol{x}$, so that $Y^{1}\mid \boldsymbol{X}{=}\boldsymbol{x}$ is equal in distribution to $G_{0}$ + $G_{1}$, where $G_{1} \independent G_{0}$ and  $G_{0}$ is equal in distribution to  $Y^{0} \mid \boldsymbol{X}{=}\boldsymbol{x}$. Then, by conditional exchangeability, $G_{0}$ is equal in distribution to $Y^{0} \mid A{=}0, \boldsymbol{X}{=}\boldsymbol{x}$ and $G_{0}+G_{1}$ is equal in distribution to $Y^{1}\mid A{=}1, \boldsymbol{X}{=}\boldsymbol{x}$. Finally, by causal consistency, $G_{0}$ is equal in distribution to $Y \mid A{=}0, \boldsymbol{X}{=}\boldsymbol{x}$ and $G_{0}+G_{1}$ is equal in distribution to $Y\mid A{=}1, \boldsymbol{X}{=}\boldsymbol{x}$. \qed

\newpage
\section{Confounder selection case study}\label{CH3CH3:A0}
As the Bayesian analysis was computationally intensive, we started by fitting a Gaussian linear mixed model (LMM) to investigate which candidate confounders did change the mean or variance of the effect of Hepatic Steatosis on the left ventricular filling pressure (LVFP). More precisely, we started by fitting \begin{equation}
Y_{i} = \beta_{0} + \boldsymbol{X}_{i}\boldsymbol{\beta_{X}}^{T} + \left(\boldsymbol{X}_{i}\boldsymbol{\beta_{\text{XA}}}^T+Z_{1i}\right)A_{i}+\varepsilon_{i},
\end{equation} where $Z_{1i}$ and $\varepsilon_{i}$ are independent and Gaussian distributed and $\mathbb{E}[\varepsilon_{i}]=0$. Initially, we include all candidate confounders: age, sex, smoke, drinks per week (DPW), diabetes state (DIAB), systolic blood pressure (SBP), antihypertensive-medication use (HRX), lipid-lowering med use (LRX), total cholesterol (CHOL), high-density lipoprotein cholesterol (HDL), triglyceride (TRIG) and fasting glucose (GLU).  

Then, we started to remove single components of $\boldsymbol{X}$ from the model to see how $\mathbb{E}[\boldsymbol{X}\boldsymbol{\beta_{\text{XA}}}^T+Z_{1}]$ was affected. If the smallest change were less than $5\%$ of $\mathbb{E}[\boldsymbol{X}\boldsymbol{\beta_{\text{XA}}}^T+Z_{1}]$, we would remove that component of $\boldsymbol{X}$ from the model. This selection procedure was continued until removing a candidate confounder from the model would result in a relative change of $\mathbb{E}[\boldsymbol{X}\boldsymbol{\beta_{\text{XA}}}^T+Z_{1}]$ greater than $5\%$. The entire procedure is shown in Table \ref{CH3tab:A0a}.

\begin{table}[h!]
	\resizebox{0.9\linewidth}{!}{
		\begin{tabular}{c||l|l|l|l|l|l|l|l|l|l|l|l|l}
			Remove fixed effect & -&age&sex&smoke&DPW&DIAB&SBP&HRX&LRX&CHOL&HDL&TRIG&GLU \\ \hline
$\mathbb{E}[\boldsymbol{X}\boldsymbol{\beta_{\text{XA}}}^T+Z_{1}]$                &0.425&0.387&0.467&0.408&0.410&0.429&0.569&0.470&0.424&0.424&0.412&0.427&0.433\\
			\textit{Relative change}     &
&-0.090&0.098&-0.040&-0.037&0.008&0.337&0.104&-0.003&-0.004&-0.033&0.004&0.018\\ \hline
$\mathbb{E}[\boldsymbol{X}\boldsymbol{\beta_{\text{XA}}}^T+Z_{1}]$                &0.424&0.383&0.466&0.407&0.409&0.428&0.568&0.471&&0.422&0.411&0.427&0.432\\
			\textit{Relative change}     &
&-0.098&0.098&-0.040&-0.037&0.008&0.338&0.111&&-0.004&-0.032&0.006&0.018\\ \hline
$\mathbb{E}[\boldsymbol{X}\boldsymbol{\beta_{\text{XA}}}^T+Z_{1}]$                &0.422&0.379&0.464&0.405&0.406&0.426&0.565&0.471&&&0.420&0.414&0.430\\
			\textit{Relative change}     &
&-0.103&0.099&-0.040&-0.038&0.009&0.337&0.115&&&-0.006&-0.019&0.017\\ \hline
$\mathbb{E}[\boldsymbol{X}\boldsymbol{\beta_{\text{XA}}}^T+Z_{1}]$                &0.420&0.362&0.415&0.414&0.421&0.427&0.558&0.473&&&&0.397&0.427\\
			\textit{Relative change}     &
&-0.139&-0.012&-0.015&0.002&0.016&0.330&0.126&&&&-0.055&0.017\\ \hline
$\mathbb{E}[\boldsymbol{X}\boldsymbol{\beta_{\text{XA}}}^T+Z_{1}]$                &0.421&0.361&0.413&0.415&&0.427&0.558&0.470&&&&0.401&0.430\\
			\textit{Relative change}     &
&-0.141&-0.019&-0.014&&0.016&0.326&0.118&&&&-0.046&0.022\\ \hline
$\mathbb{E}[\boldsymbol{X}\boldsymbol{\beta_{\text{XA}}}^T+Z_{1}]$                &0.415&0.361&0.406&&&0.422&0.550&0.468&&&&0.402&0.424\\
			\textit{Relative change}     &
&-0.129&-0.022&&&0.016&0.326&0.128&&&&-0.032&0.023\\ \hline
$\mathbb{E}[\boldsymbol{X}\boldsymbol{\beta_{\text{XA}}}^T+Z_{1}]$                &0.422&0.368&0.414&&&&0.556&0.482&&&&0.407&0.462\\
			\textit{Relative change}     &
&-0.126&-0.018&&&&0.319&0.143&&&&-0.036&0.096\\ \hline
$\mathbb{E}[\boldsymbol{X}\boldsymbol{\beta_{\text{XA}}}^T+Z_{1}]$                &0.414&0.350&&&&&0.547&0.473&&&&0.368&0.449\\
			\textit{Relative change}     &
&-0.154&&&&&0.321&0.142&&&&-0.112&0.084\\
	\end{tabular}}
 \centering
	\caption{Change in the estimate of exposure effect ($\mathbb{E}[\boldsymbol{X}\boldsymbol{\beta_{\text{XA}}}^T+Z_{1}]$) after removing candidate confounders from the Gaussian LMM. In each step of the procedure, the covariate with the least impact is removed. }\label{CH3tab:A0a}
\end{table} \noindent At this stage age, SBP, HRX, TRIG, and GLU affect $\mathbb{E}[\boldsymbol{X}\boldsymbol{\beta_{\text{XA}}}^T+Z_{1}]$ and are therefore considered confounders. The estimates of this model can be found in Table \ref{CH3tab:A0b}. \begin{table}[h!]
	\begin{tabular}{ll|ll}
		\multicolumn{2}{c|}{\textbf{Fixed effects}} & \multicolumn{2}{c}{\textbf{Variance components}} \\ \hline
		Intercept&6.330& & \\
Hepatic Steatosis&0.414& Hepatic Steatosis& 1.805 \\
age&0.484& & \\
Hepatic Steatosis*age&0.003& & \\
SBP&0.238& & \\
Hepatic Steatosis*SBP&0.242& & \\
HRX&0.392& & \\
Hepatic Steatosis*hrxS&0.071& & \\
TRIG&0.034& & \\
Hepatic Steatosis*TRIG&-0.128& & \\
GLU&0.006& & \\
Hepatic Steatosis*GLU&0.146& & \\
 & & Residual & 2.538       \\     
	\end{tabular} \centering
	\caption{Gaussian LMM parameter estimates ignoring confounder effect heterogeneity. }\label{CH3tab:A0b}
\end{table} However, candidate confounders should also be recognized as confounders when they do not seriously affect the mean $\mathbb{E}[\boldsymbol{X}\boldsymbol{\beta_{\text{XA}}}^T+Z_{1}]$ but instead affect the variance of $Z_{1}$. First, we add a random effect for dichotomized confounders selected at this stage and fit  \begin{equation}
\begin{cases}
Y_{i} = \beta_{0} + \boldsymbol{X}_{i}\boldsymbol{\beta_{X}}^{T} + \left(\boldsymbol{X}\boldsymbol{\beta_{\text{XA}}}^T+Z_{1}\right)A_{i}+\delta_{i},\\
\delta_{i}=\varepsilon_{i}+\boldsymbol{\tilde{X}}_{i}\boldsymbol{Z}_{Xi}^{T}
\end{cases}
\end{equation} where $Z_{1i}$, $\varepsilon_{i}$ and $\boldsymbol{Z}_{Xi}$ are independent and Gaussian distributed and $\mathbb{E}[\varepsilon_{i}]=0$. Dichotomized confounders, $\boldsymbol{\tilde{X}}_{i}$ are obtained by comparing the level of the continuous confounder with the sample median; $>$ when the sample variance is larger in individuals with higher values of the outcome and $<$ otherwise. The estimates can be found in Table \ref{CH3tab:A0c}, and it becomes clear that the variance of $Z_{1}$ seriously changes after accounting for possible confounding-effect heterogeneity. 

\begin{table}[h!]
	\begin{tabular}{ll|ll}
		\multicolumn{2}{c|}{\textbf{Fixed effects}} & \multicolumn{2}{c}{\textbf{Variance components}} \\ \hline
		Intercept&6.335&& \\
Hepatic Steatosis&0.408&Hepatic Steatosis&0.940 \\
age&0.476&(age > 49)&1.069 \\
Hepatic Steatosis*age&0.003&& \\
SBP&0.248&(SBP > 120)&0.503 \\
Hepatic Steatosis*SBP&0.215&& \\
HRX&0.402&HRX&1.489 \\
Hepatic Steatosis*hrxS&0.090&& \\
TRIG&0.027&(TRIG > 98)&0.051 \\
Hepatic Steatosis*TRIG&-0.082&& \\
GLU&-0.009&(GLU < 96)&0.194 \\
Hepatic Steatosis*GLUCOSE&0.138&& \\
&&Residual&1.441\\
	\end{tabular}
 \centering
	\caption{Gaussian LMM parameter estimates for the model after first selection procedure accounting for confounder effect heterogeneity.}\label{CH3tab:A0c}
\end{table}

Subsequently, we again added one of the other candidate confounders to the model and investigated how the variance of $Z_{1}$ changed. If the relative variance change was more than $10\%$ for one of the covariates, we added the most influential candidate confounder to the model. After adding the covariate sex and diabetes state to the model, no other candidate confounders gave rise to a change in variance greater than $10\%$ as shown in Table \ref{CH3tab:A0d}. 

\begin{table}[h!]
\resizebox{\linewidth}{!}{
		\begin{tabular}{l|l|l|l|l|l|l|l|l}
			Add random effect & -     & sex   & smoke  & (DPW$<1.5$) & DIAB & (CHOL > 189)  & (HDL > 52)   & 1-LRX  \\ \hline
			Variance $Z_{1}$  &    0.940&0.712&0.841&0.945&0.873&0.958&0.945&0.979\\
\textit{Relative change}&&0.243&0.105&0.005&-0.072&0.019&0.006&0.042\\ \hline
Variance $Z_{1}$  & 0.712&&0.689&0.731&0.589&0.759&0.719&0.743\\
\textit{Relative change}&&&-0.032&0.027&-0.173&0.066&0.010&0.044\\ \hline
Variance $Z_{1}$  & 0.589&&0.552&0.635&&0.636&0.593&0.623\\
\textit{Relative change}&&&-0.063&0.078&&0.080&0.007&0.057\\
	\end{tabular}}
 \centering
	\caption{Change in the estimate of the variance of the random effect of fatty liver (variance of $Z_{1}$) after adding candidate confounders to the LMM. In each step of the procedure, the covariate with the most impact is added. }\label{CH3tab:A0d}
\end{table}

As a final step, we check whether we can remove any of the candidate confounders from the model without significantly changing the variance of $Z_{1}$ (again 10$\%$) or $\mathbb{E}[\boldsymbol{X}\boldsymbol{\beta_{\text{XA}}}^T+Z_{1}]$ (again 5$\%$). 

\begin{table}[h!]
		\begin{tabular}{l|l|l|l|l|l|l|l|l}
Remove&-&age&sex&DIAB&SBP&HRX&TRIG&GLU\\ \hline
Variance $Z_{1}$&0.589&0.423&0.830&0.712&0.886&0.956&0.612&0.595\\
$\mathbb{E}[\boldsymbol{X}\boldsymbol{\beta_{\text{XA}}}^T+Z_{1}]$&0.378&0.309&0.392&0.393&0.521&0.459&0.390&0.387\\ \hline
Variance $Z_{1}$&0.595&0.465&0.824&0.745&0.886&0.972&0.627&\\
$\mathbb{E}[\boldsymbol{X}\boldsymbol{\beta_{\text{XA}}}^T+Z_{1}]$&0.387&0.326&0.395&0.432&0.537&0.467&0.405&\\ \hline
Variance $Z_{1}$&0.627&0.525&0.813&0.776&0.934&0.997&&\\
$\mathbb{E}[\boldsymbol{X}\boldsymbol{\beta_{\text{XA}}}^T+Z_{1}]$&0.405&0.334&0.372&0.458&0.567&0.471&&\\
	\end{tabular}
 \centering
	\caption{Change in the estimated mean and the variance of the random effect of fatty liver after removing candidate confounders from the LMM.}\label{CH3tab:A0var}
\end{table}

With this procedure, we have selected age, sex, DIAB, SBP, and HRX as confounders. The parameter estimates of the final Gaussian LMM are presented in Table \ref{CH3tab:A0e} and were used as initial values in the Bayesian procedure of the case study. 
\begin{table}[t]
	\begin{tabular}{ll|ll}
		\multicolumn{2}{c|}{\textbf{Fixed effects}} & \multicolumn{2}{c}{\textbf{Variance components}} \\ \hline
		Intercept&5.882&&\\
Hepatic Steatosis&0.319&Hepatic Steatosis&0.627\\
age&0.366&(age > 49)&0.706\\
Hepatic Steatosis*age&0.007&&\\
SBP&0.311&(SBP > 120)&0.724\\
Hepatic Steatosis*SBP&0.174&&\\
HRX&0.418&HRX&1.370\\
Hepatic Steatosis*HRX&0.057&&\\
SEX&0.650&SEX&0.927\\
Hepatic Steatosis*SEX&0.118&&\\
DIAB&0.471&DIAB&0.775\\
Hepatic Steatosis*DIAB&0.197&&\\
&&Residual&1.063\\      
	\end{tabular}
 \centering
	\caption{Final LMM parameter estimates accounting for age, sex, diabetes, systolic blood pressure, and antihypertensive medication use. }\label{CH3tab:A0e}
\end{table}
\newpage
\section{Supplementary figures} \vspace{-1cm}
\begin{figure}[h!]
	\centering
	\begin{subfigure}{.325\textwidth}
		\resizebox{1\linewidth}{!}{\includegraphics{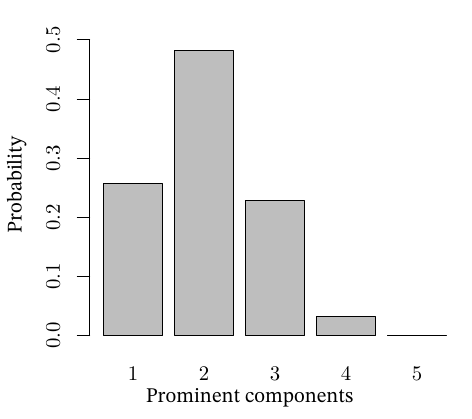}}
		\centering\caption{}\label{CH3DPa}	
	\end{subfigure}
	\begin{subfigure}{.325\textwidth}
		\resizebox{\linewidth}{!}{\includegraphics{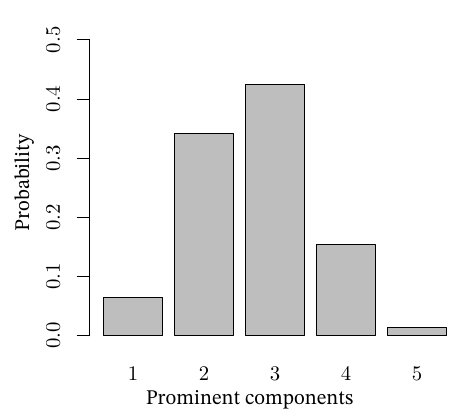}}
		\centering\caption{}\label{CH3DPb}	
	\end{subfigure}
	\begin{subfigure}{.325\textwidth}
		\resizebox{1\linewidth}{!}{\includegraphics{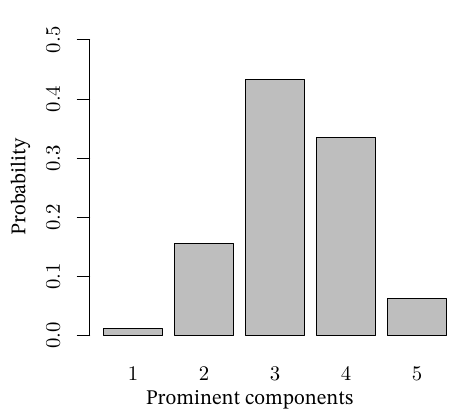}}
		\centering\caption{}\label{CH3DPc}	
	\end{subfigure}
	\caption{Distribution of the number of $p_{j} \in \boldsymbol{p}{:}~p_{j}{>}0.10$ when $\boldsymbol{p} ~{\sim}~ \text{Dir}(\alpha,\alpha,\alpha,\alpha,\alpha)$ for $\alpha=\frac{1}{5}$ (a), $\frac{1}{2}$ (b) and $1$ (c).}\label{CH3Dprior}	
\end{figure}\vspace{-1cm}
\begin{figure}[h!]
	\centering
	\begin{subfigure}{.325\textwidth}
		\resizebox{1\linewidth}{!}{\includegraphics{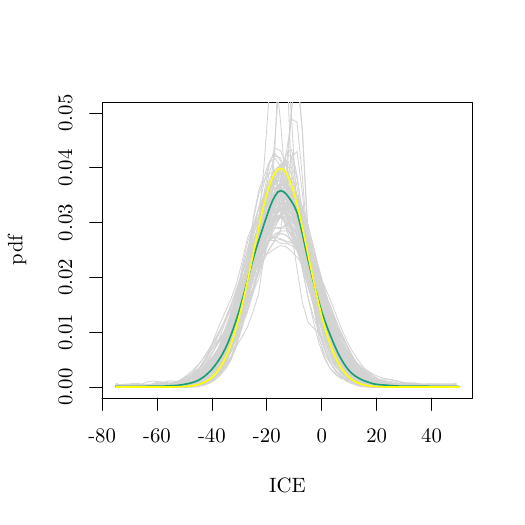}}
		\caption{}\label{CH33a}	
	\end{subfigure}
	\begin{subfigure}{.325\textwidth}
		\resizebox{1\linewidth}{!}{\includegraphics{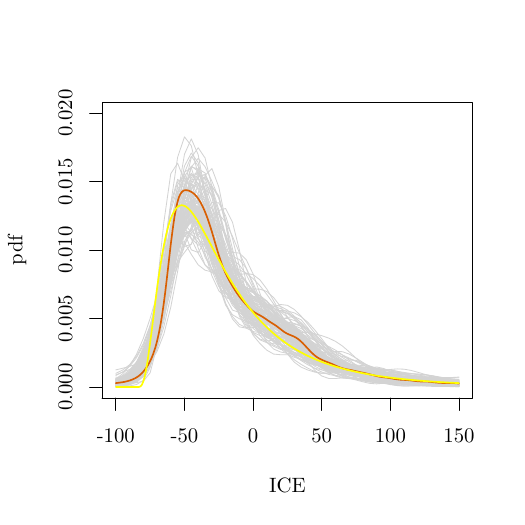}}
		\caption{}\label{CH33b}	
	\end{subfigure}
	\begin{subfigure}{.325\textwidth}
		\resizebox{1\linewidth}{!}{\includegraphics{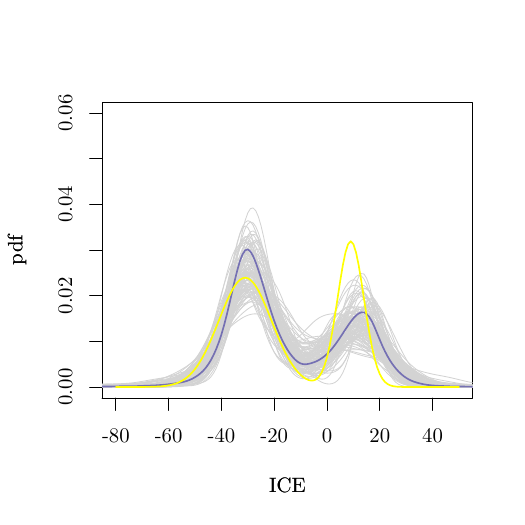}}
		\caption{}\label{CH33c}	
	\end{subfigure}
	\caption{Kernel density estimates of the $Z_{1}$ sampled per MCMC iteration (grey) for one of the datasets used in Figure \ref{CH3FIG3}. The corresponding posterior distributions (green/orange/blue) and the actual ICE distribution (yellow) are also presented.} \label{CH3FIG3b}
\end{figure} \vspace{-2cm}
\begin{figure}[h!]
	\centering
	\includegraphics[width=0.95\linewidth]{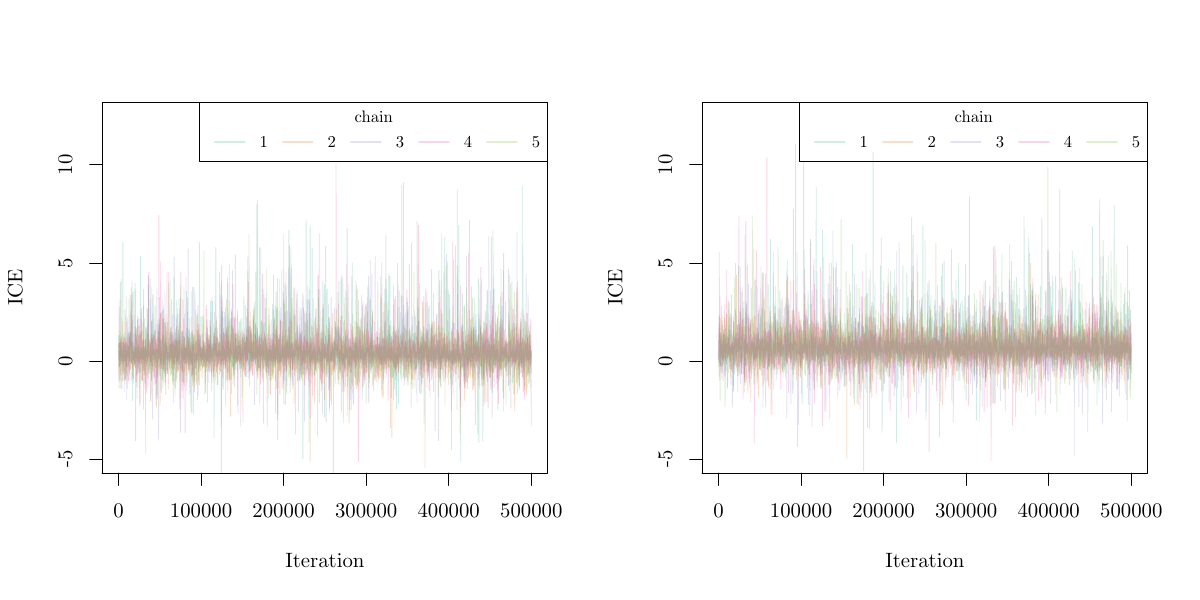} 
	\vspace{-0.5cm}\caption{The trace plots of simulated ICEs for two unexposed individuals while fitting model \eqref{CH3LVmodel3}.}
	\label{CH3fig:traceplot}
\end{figure}

\newpage 
\subsection{Posterior predictive distribution per confounder strata} \label{CH3appNOC}
\begin{figure}[h!]
	\centering
	\includegraphics[width=0.75\linewidth]{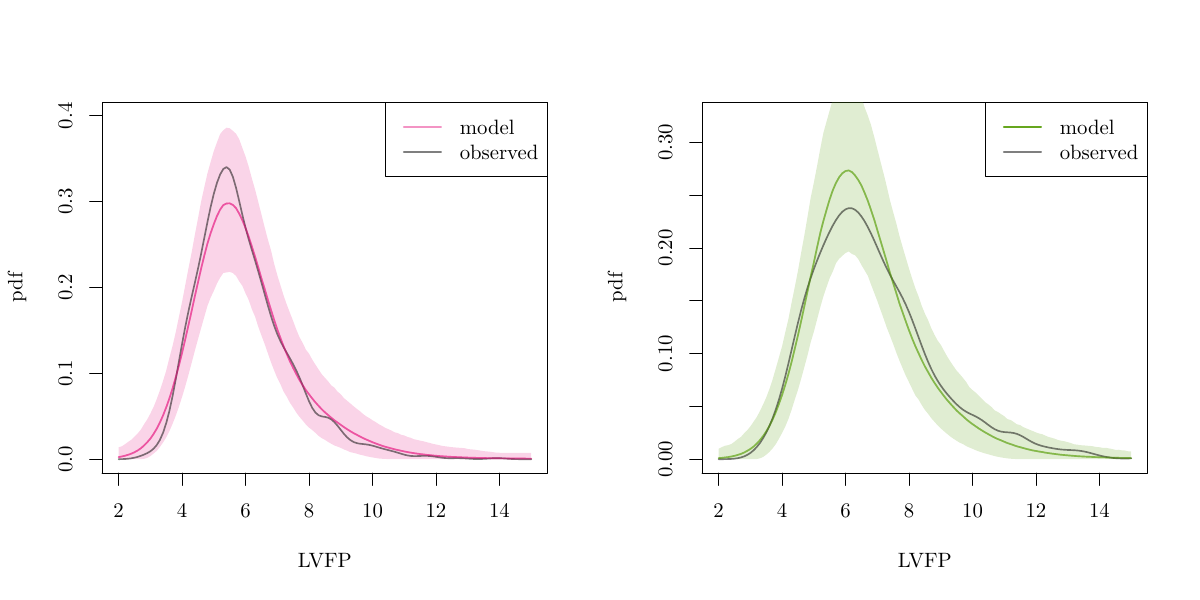} 
	\vspace{-0.25cm}\caption{Posterior predictive distribution of the LVFP for the model \eqref{CH3LVmodel3}, with corresponding $95 \%$ Bayesian credible intervals and the kernel density of the observed data for individuals that are older than 49 years (right) or not (left). }\label{CH3fig:ageNOC}
\end{figure}
\vspace{-1cm}
\begin{figure}[h!]
	\centering
	\includegraphics[width=0.75\linewidth]{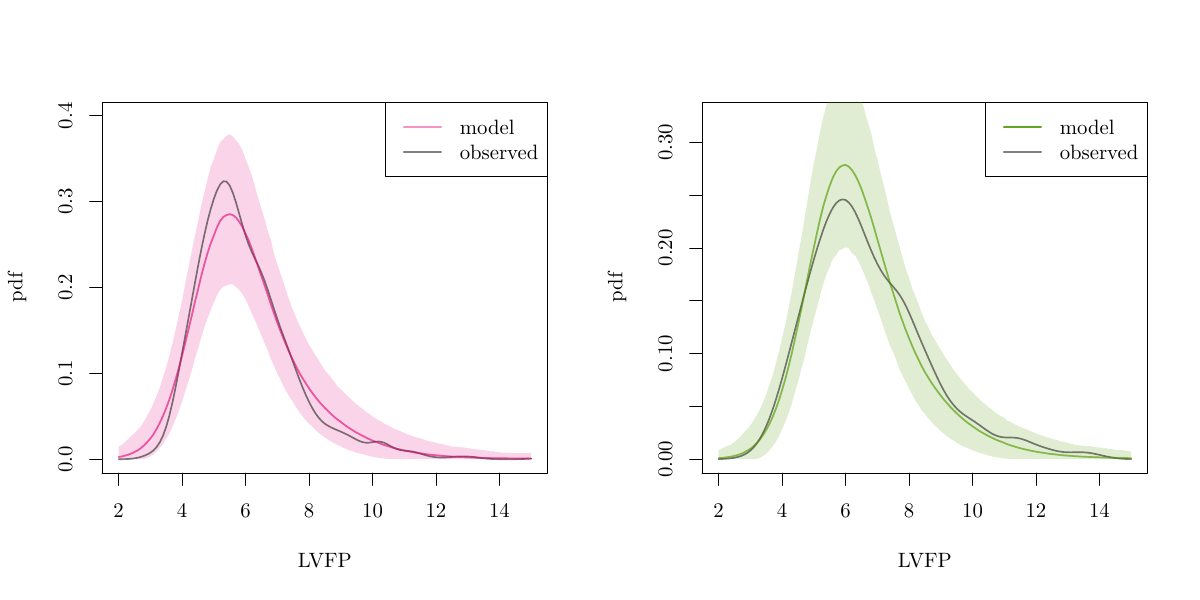}
	\vspace{-0.25cm}\caption{Posterior predictive distribution of the LVFP for the model \eqref{CH3LVmodel3}, with corresponding $95 \%$ Bayesian credible intervals and the kernel density of the observed data for males (left) and females (right). }\label{CH3fig:sexNOC}
\end{figure} 
\vspace{-1cm}
\begin{figure}[h!]
	\centering
	\includegraphics[width=0.75\linewidth]{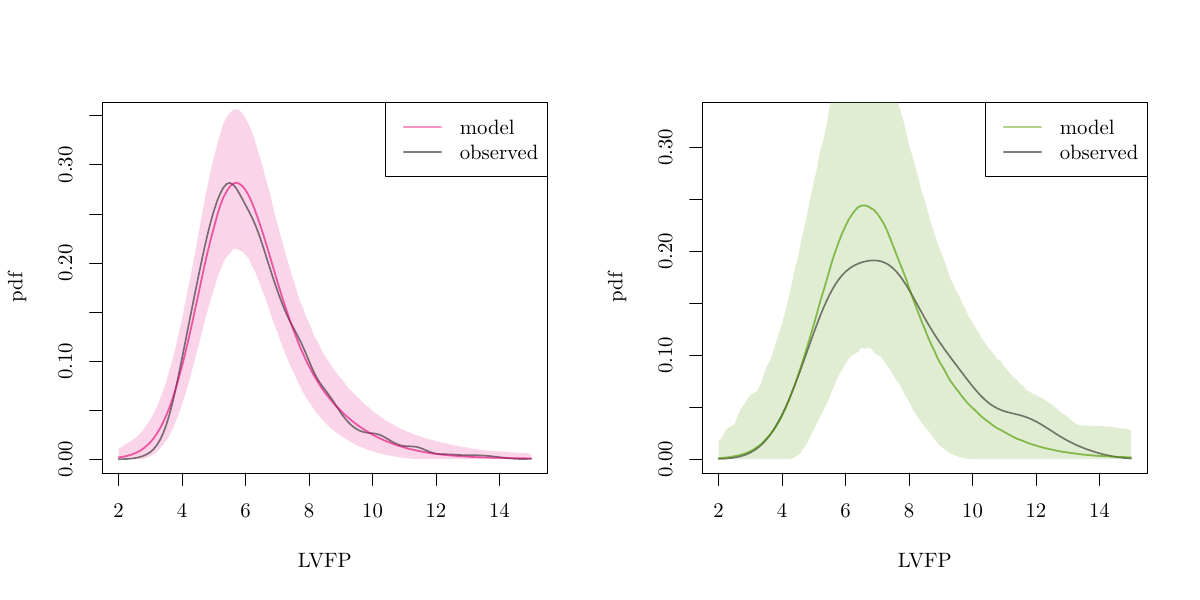}
	\vspace{-0.25cm}\caption{Posterior predictive distribution of the LVFP for the model \eqref{CH3LVmodel3}, with corresponding $95 \%$ Bayesian credible intervals and the kernel density of the observed data for individuals with (right) and without (left) diabetes. }\label{CH3fig:diabNOC}
\end{figure}
\begin{figure}[h!]
	\centering
	\includegraphics[width=0.75\linewidth]{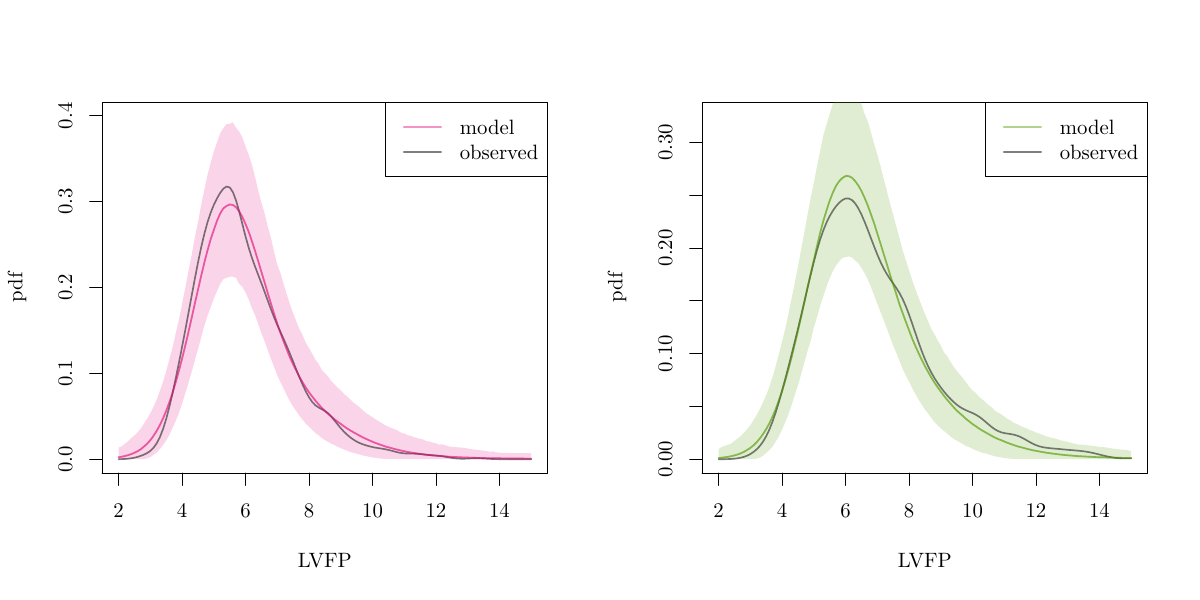}
	\vspace{-0.25cm}\caption{Posterior predictive distribution of the LVFP for the model \eqref{CH3LVmodel3}, with corresponding $95 \%$ Bayesian credible intervals and the kernel density of the observed data for individuals with a systolic blood pressure above $120$ mmHg (right) or not (left). }\label{CH3fig:sbpNOC}
\end{figure} 
\begin{figure}[h!]
	\centering
	\includegraphics[width=0.75\linewidth]{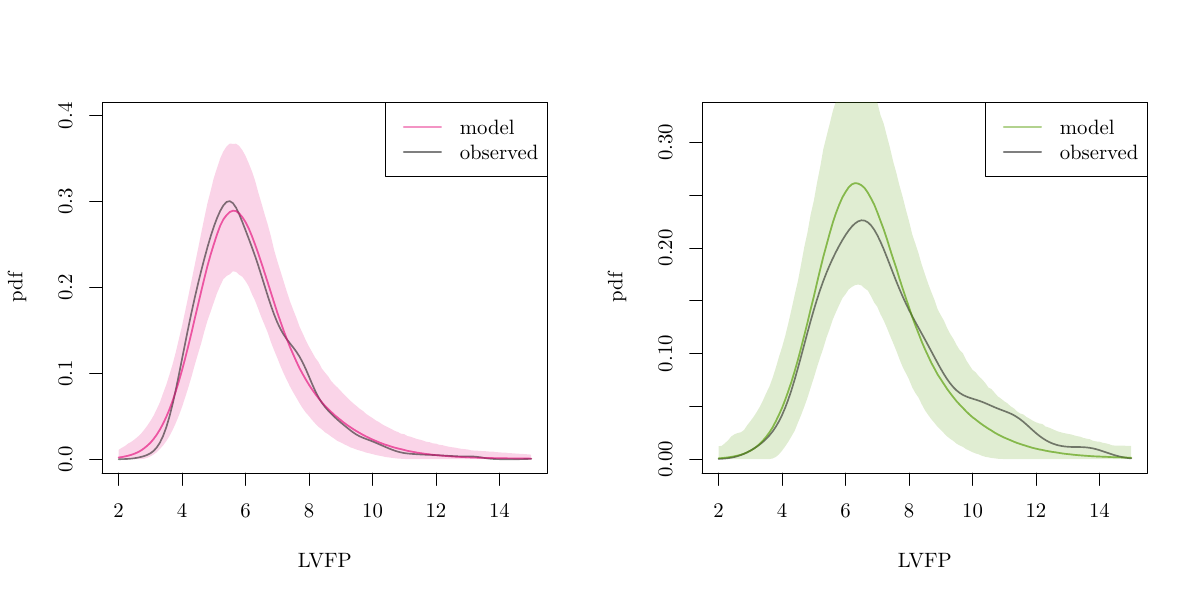}
	\vspace{-0.25cm}\caption{Posterior predictive distribution of the LVFP for the model \eqref{CH3LVmodel3}, with corresponding $95 \%$ Bayesian credible intervals and the kernel density of the observed data for individuals with (right) and without (left) antihypertensive-medication use. }\label{CH3fig:hrxNOC}
\end{figure}
\newpage $ $
\newpage
\subsection{Model adjusting for confounding-effect heterogeneity}\label{CH3appCH} \vspace{-1cm}
\begin{figure}[h!]
	\centering
	\includegraphics[width=0.75\linewidth]{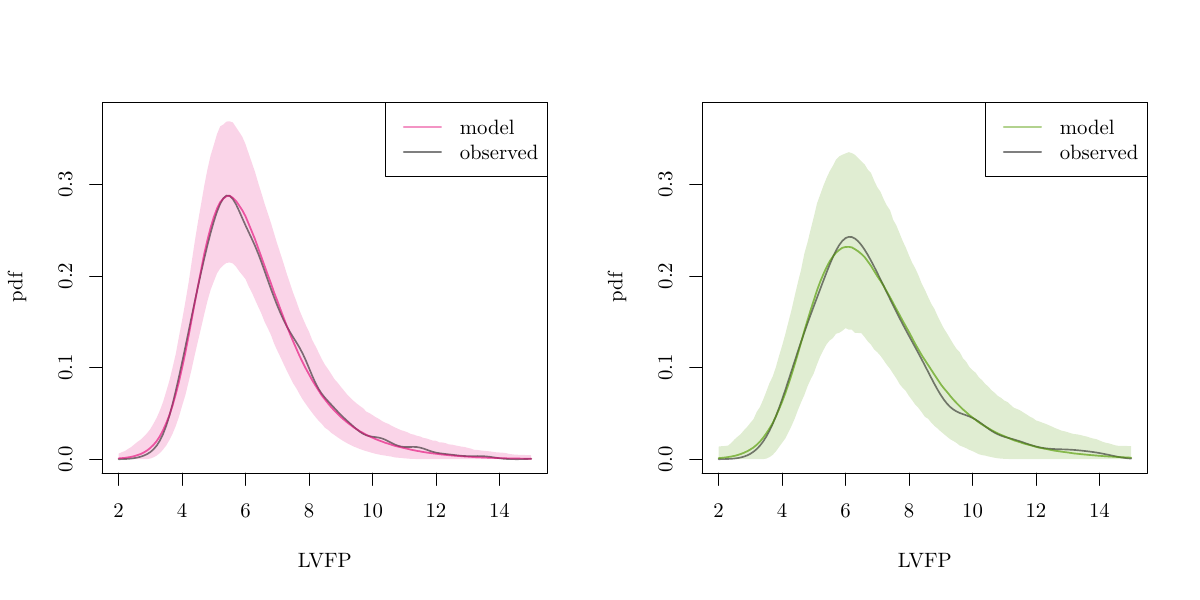}
	\caption{Posterior predictive distribution of the LVFP for model   \eqref{CH3LVmodel2}, with corresponding $95 \%$ Bayesian credible intervals and the kernel density of the observed data for unexposed (left) and exposed individuals (right). }
	\label{CH3fig:fattyliver}
\end{figure}

\vspace{-1cm}
\begin{figure}[h!]
	\centering
	\includegraphics[width=0.75\linewidth]{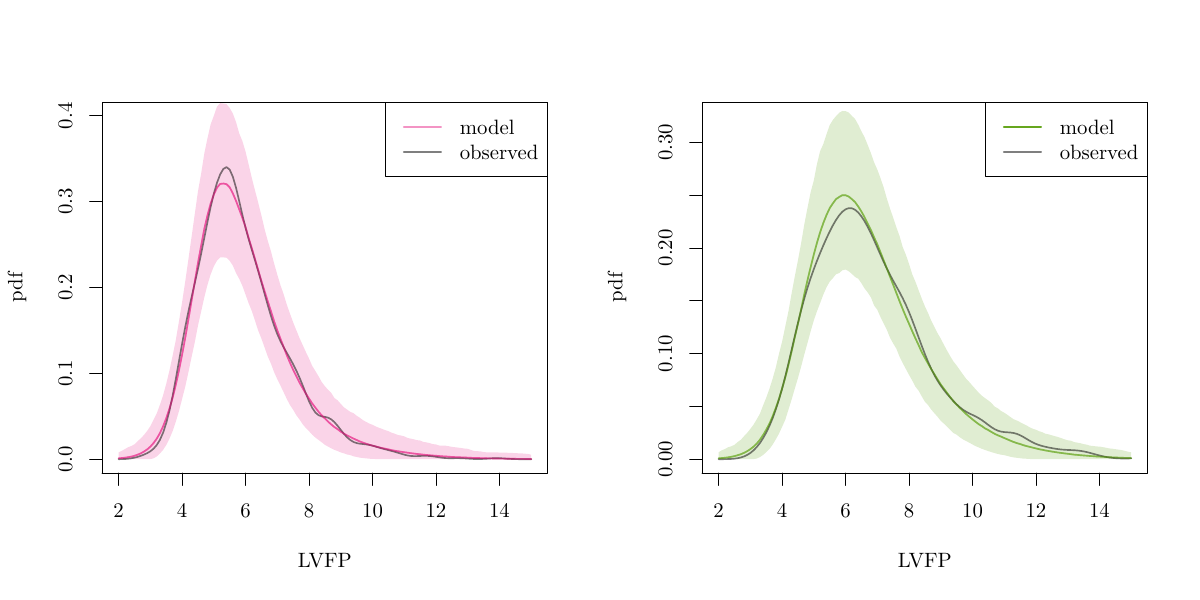}
	\vspace{-0.25cm}\caption{Posterior predictive distribution of the LVFP for the model \eqref{CH3LVmodel2}, with corresponding $95 \%$ Bayesian credible intervals and the kernel density of the observed data for individuals that are older than 49 years (right) or not (left). }
	\label{CH3fig:age}
\end{figure}\vspace{-1cm}
\begin{figure}[h!]
	\centering
	\includegraphics[width=0.75\linewidth]{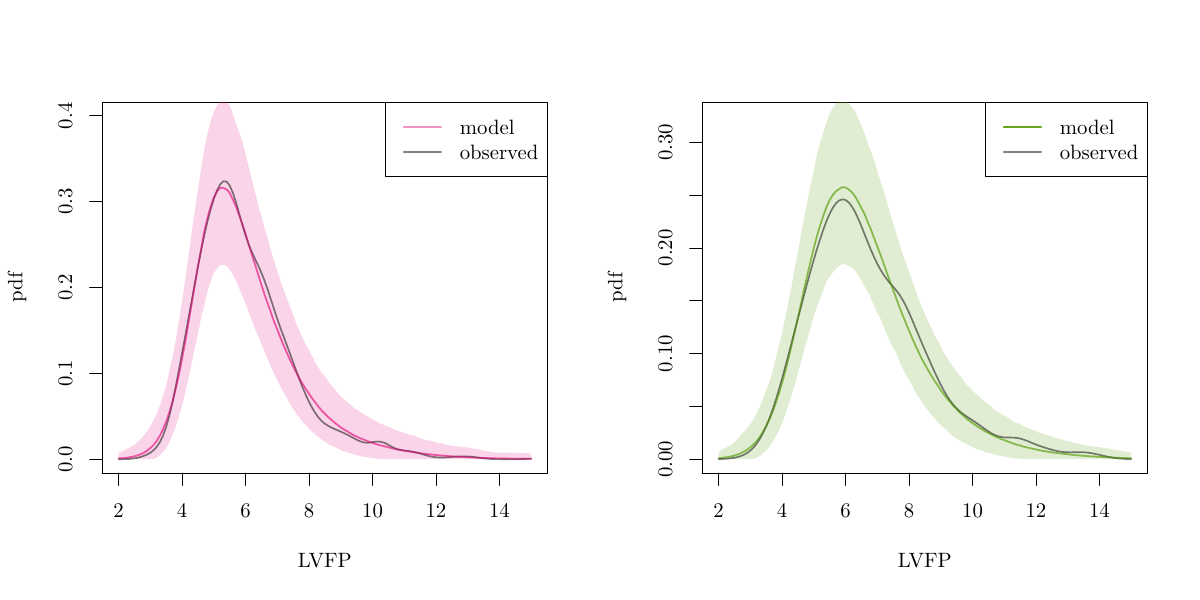}
	\vspace{-0.25cm}\caption{Posterior predictive distribution of the LVFP for the model \eqref{CH3LVmodel2}, with corresponding $95 \%$ Bayesian credible intervals and the kernel density of the observed data for males (left) and females (right). }\label{CH3fig:sex}
\end{figure} 
\vspace{-1cm}
\begin{figure}[h!]
	\centering
	\includegraphics[width=0.75\linewidth]{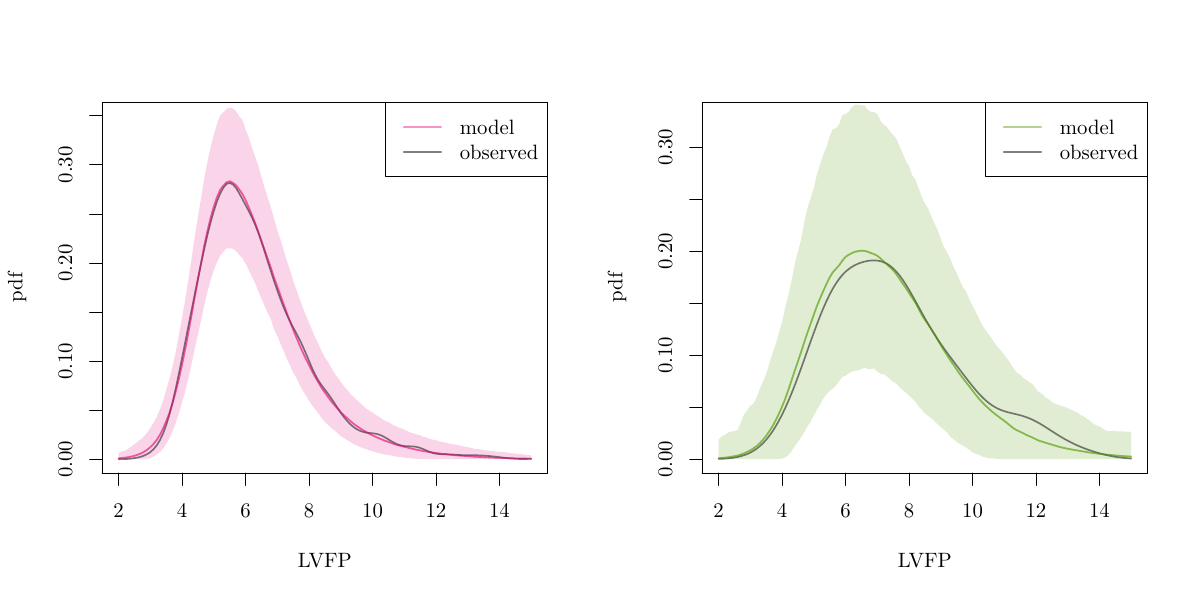}
	\vspace{-0.25cm}\caption{Posterior predictive distribution of the LVFP for the model \eqref{CH3LVmodel2}, with corresponding $95 \%$ Bayesian credible intervals and the kernel density of the observed data for individuals with (right) and without (left) diabetes. } \label{CH3fig:diab}
\end{figure}

\begin{figure}[h!]
	\centering
	\includegraphics[width=0.75\linewidth]{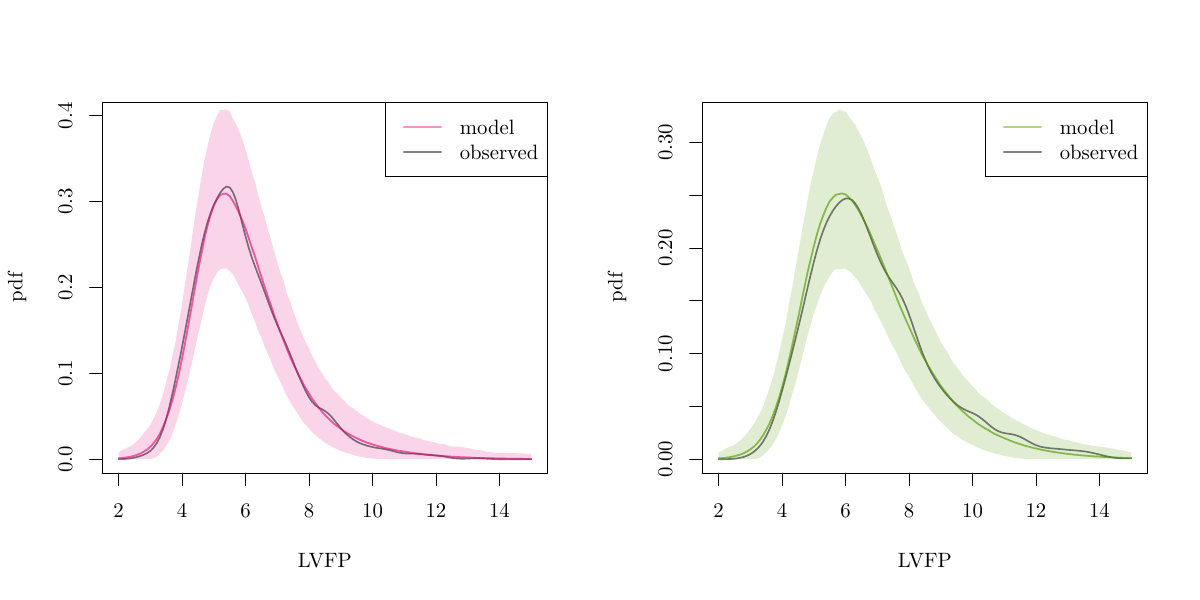}
	\vspace{-0.25cm}\caption{Posterior predictive distribution of the LVFP for the model \eqref{CH3LVmodel2}, with corresponding $95 \%$ Bayesian credible intervals and the kernel density of the observed data for individuals with a systolic blood pressure above $120$ mmHg (right) or not (left). }\label{CH3fig:sbp}
\end{figure} \begin{figure}[h]
	\centering
	\includegraphics[width=0.75\linewidth]{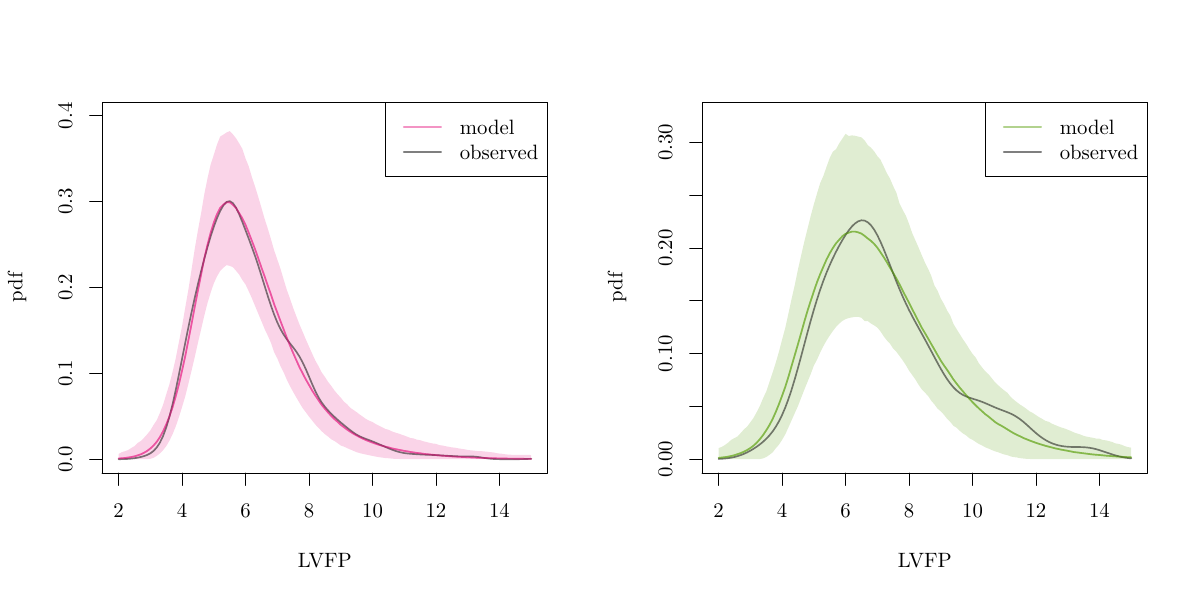}
	\vspace{-0.25cm}\caption{Posterior predictive distribution of the LVFP for the model with \eqref{CH3LVmodel2}, with corresponding $95 \%$ Bayesian credible intervals and the kernel density of the observed data for individuals with (right) and without (left) antihypertensive-medication use. }\label{CH3fig:hrx}
\end{figure}
\newpage $ $
\newpage $ $
\newpage
\subsection{Model with Gaussian residual}\label{CH3appNOFR}
\begin{figure}[h!]
	\centering
	\includegraphics[width=0.75\linewidth]{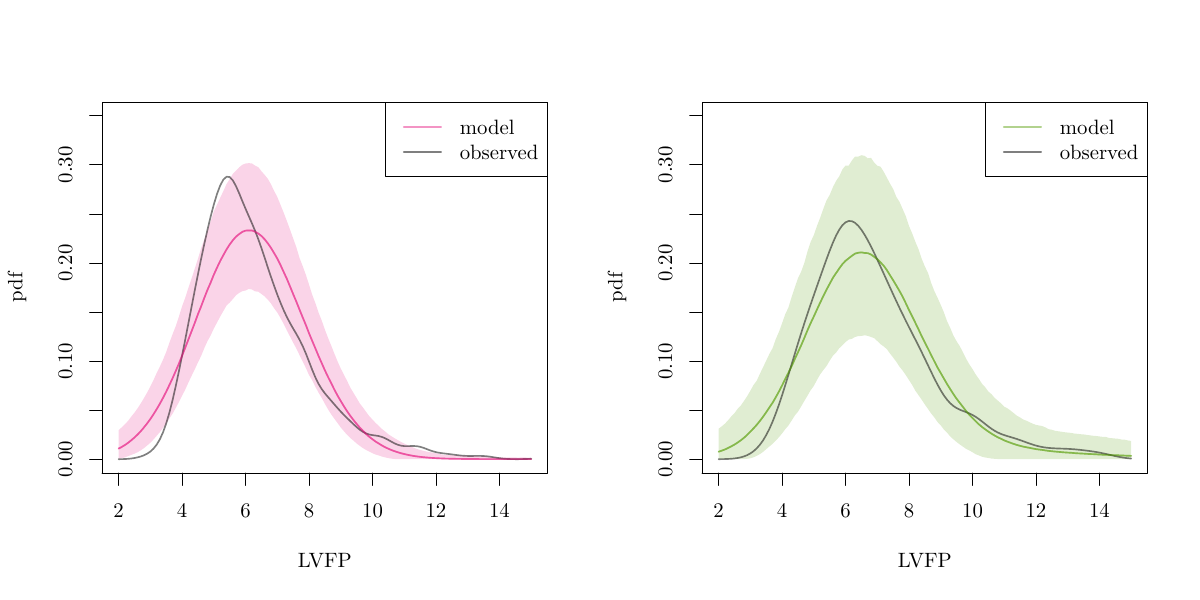}
	\caption{Posterior predictive distribution of the LVFP for model   \eqref{CH3LVmodel3} but with Gaussian distributed $\varepsilon$, with corresponding $95 \%$ Bayesian credible intervals and the kernel density of the observed data for unexposed (left) and exposed individuals (right). }
	\label{CH3fig:fattyliverNOFR}
\end{figure}

\subsection{Gaussian LMM}\label{CH3appLMM} 
\begin{figure}[h!]
	\centering
	\includegraphics[width=0.75\linewidth]{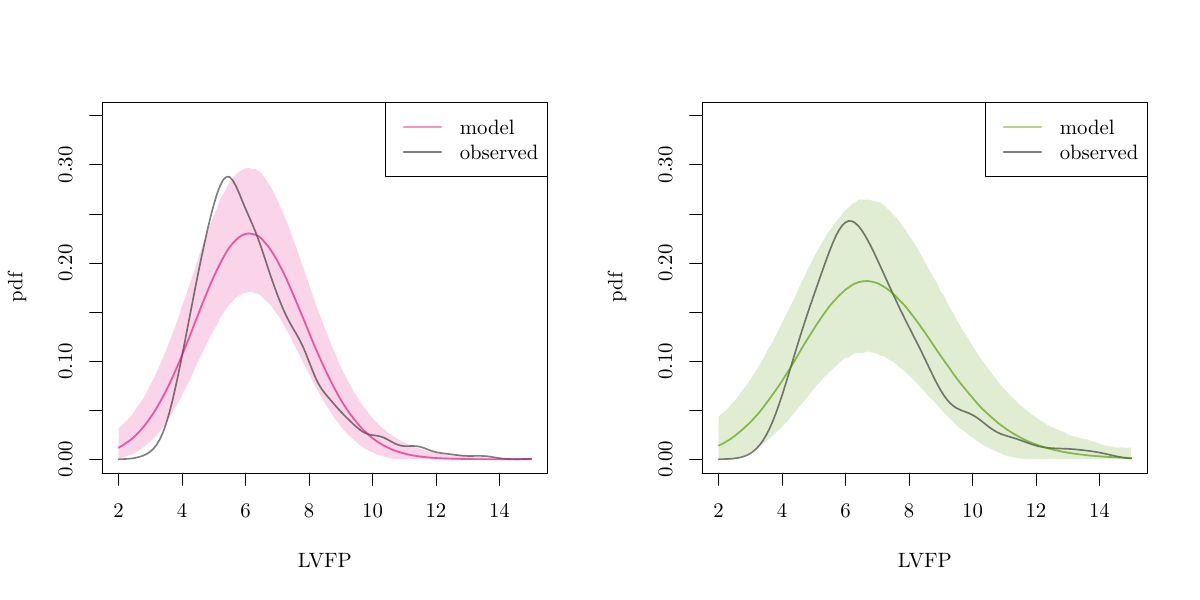}
	\caption{Posterior predictive distribution of the LVFP for model   \eqref{CH3LVmodel3} but with Gaussian distributed $Z_{1}$ and $\varepsilon$, with corresponding $95 \%$ Bayesian credible intervals and the kernel density of the observed data for unexposed (left) and exposed individuals (right). }
	\label{CH3fig:fattyliverLMM}
\end{figure}
\vfill
\end{document}